\documentclass[structabstract]{aa}  
\usepackage{graphicx}
\usepackage{textcomp}
\usepackage{lscape}
\usepackage[varg]{txfonts}
\usepackage{mathrsfs}
\usepackage{amssymb}
\usepackage{natbib}
\usepackage[colorlinks, citecolor={blue}]{hyperref}

\begin{document}

   \title{Catalogues of isolated galaxies, isolated pairs, and isolated triplets in the local Universe\thanks{Tables 1, 2, 3, 5, and 6 are only available in electronic form at the CDS via anonymous ftp to cdsarc.u-strasbg.fr (130.79.128.5) or via \texttt{http://cdsweb.u-strasbg.fr/cgi-bin/qcat?J/A+A/}.}}

   \author{M. Argudo-Fern\'andez\inst{1,2,3}
          \and
          S. Verley\inst{1,4}
          \and
          G. Bergond\inst{5}
          \and
          S. Duarte Puertas\inst{1,2}
          \and
          E. Ramos Carmona\inst{1}
          \and
          J. Sabater\inst{6}
          \and \\
          M. Fern\'andez Lorenzo\inst{2}         
          \and
          D. Espada\inst{7,8,9}
          \and          
          J. Sulentic \inst{2}
          \and
          J.\,E.\,Ruiz\inst{2}
          \and
          S.\,Leon\inst{7}}

   \institute{Departamento de F\'isica Te\'orica y del Cosmos, Universidad de Granada, 18071 Granada, Spain
         \and
             Instituto de Astrof\'isica de Andaluc\'ia (CSIC) Apdo. 3004, 18080 Granada, Spain
         \and
             Key Laboratory for Research in Galaxies and Cosmology, Shanghai Astronomical Observatory, Chinese Academy of Sciences, 80 Nandan Road, Shanghai, China, 200030
         \and
             Instituto Universitario Carlos I de F\'isica Te\'orica y Computacional, Universidad de Granada, 18071 Granada, Spain
         \and
             Centro Astron\'omico Hispano Alem\'an, Compl. Observatorio Calar Alto s/n, Sierra de los Filabres, 04550, Gerg\'al, Spain   
         \and
             Institute for Astronomy, University of Edinburgh, Edinburgh EH9 3HJ, UK
         \and
             Joint ALMA Observatory (ALMA/ESO), Alonso de C\'ordova 3107, Vitacura, Santiago 763-0355, Chile 
         \and
             National Astronomical Observatory of Japan (NAOJ), 2-21-1 Osawa, Mitaka, Tokyo 181-8588, Japan
         \and      
             Department of Astronomical Science, The Graduate University for Advanced Studies (SOKENDAI), 2-21-1 Osawa, Mitaka, Tokyo 181-8588, Japan 
             }

   \date{Received 4 May 2015; accepted 31 May 2015}

 
\abstract
{The construction of catalogues of galaxies and the a posteriori study of galaxy properties in relation to their environment have been hampered by scarce redshift information. The new 3-dimensional (3D) surveys permit small, faint, physically bound satellites to be distinguished from a background-projected galaxy population, giving a more comprehensive 3D picture of the surroundings.}
{We aim to provide representative samples of isolated galaxies, isolated pairs, and isolated triplets for testing galaxy evolution and secular processes in low density regions of the local Universe, as well as to characterise their local and large-scale environments.}
{We used spectroscopic data from the tenth data release of the Sloan Digital Sky Survey (SDSS-DR10) to automatically and homogeneously compile catalogues of 3~702 isolated galaxies, 1~240 isolated pairs, and 315 isolated triplets in the local Universe ($z\leq0.080$). To quantify the effects of their local and large-scale environments, we computed the projected density and the tidal strength for the brightest galaxy in each sample.}
{We find evidence of isolated pairs and isolated triplets that are physically bound at projected separations up to $d~\leq~450$\,kpc with radial velocity difference $\Delta\,\varv~\leq~160$\,km\,s$^{-1}$, where the effect of the companion typically accounts for more than 98\% of the total tidal strength affecting the central galaxy. For galaxies in the catalogues, we provide their positions, redshifts, and degrees of relation with their physical and large-scale environments. The catalogues are publicly available to the scientific community.}
{For isolated galaxies, isolated pairs, and isolated triplets, there is no difference in their degree of interaction with the large-scale structure (up to 5\,Mpc), which may suggest that they have a common origin in their formation and evolution. We find that most of them belong to the outer parts of filaments, walls, and clusters, and generally differ from the void population of galaxies.}

   \keywords{galaxies: general  --
             galaxies: formation  --
             galaxies: evolution}

   \maketitle
%

\section{Introduction}  \label{Sec:intro3}
   
It has long been known that galaxies are not distributed uniformly through space \citep{1958ApJS....3..211A,1984ApJ...281...95P,1987MNRAS.226..543E}. Recent galaxy redshift surveys like the Sloan Digital Sky Survey \citep[SDSS;][]{2000AJ....120.1579Y,2011AJ....142...72E}, which map the large-scale structures (LSS) of the Universe in three dimensions (3D), have confirmed that galaxies are distributed in a hierarchical structure of filaments and walls, which surround large voids \citep{2007ApJ...658..884C,2012MNRAS.426.3041H, 2012MNRAS.421..926P}.  More than half of the galaxies (54\%) in the local Universe ($z~\lesssim~0.1$) appear concentrated in virialised groups and clusters \citep{2013AJ....146...69C}. Another 20\% of galaxies are located in collapsing regions around groups and clusters \citep{1987ApJ...321..280T,2007ApJ...655..790C,2011MNRAS.412.2498M}. The remaining populations of galaxies (26\%) are referred to as ``field'' galaxies. A small fraction of these galaxies can be found in loose pairs and compact groups 
\citep{1991ApJ...374..407X,2011AstBu..66..389K}.

The fraction of isolated galaxies \citep[galaxies that have not been appreciably affected by their closest neighbours during a past crossing time $t_{\rm{cc}} \approx$\,3\,Gyr,][]{2005A&A...436..443V} is still uncertain. Short lists of isolated galaxies began to appear after it became clear that interactions produced significant radio and optical effects on galaxies \citep{1976ApJS...32..171S,1978ApJ...219...46L}. The first major compilation of isolated galaxies, the \textbf{C}atalogue of \textbf{I}solated \textbf{G}alaxies \citep[CIG,][]{1973SoSAO...8....3K}, involved about 1000 objects visually selected from the First Palomar Observatory Sky Survey. Studies of this sample, carried out by the AMIGA (\textbf{A}nalysis of the interstellar \textbf{M}edium of \textbf{I}solated \textbf{GA}laxies\footnote{\texttt{http://amiga.iaa.es}}) project, represented the first statistically significant investigations, while others focussed more on detailed studies of smaller samples \citep{1998MNRAS.295...99M,1999A&A...344..421M,2004A&A...420..873V}. 

With the advent of new digital data, the estimation of the fraction of isolated galaxies has been a topic of growing interest \citep{2005AJ....129.2062A,2009MNRAS.394.1409E,2010ASPC..421...69K,2010Ap.....53..462K,2010AstBu..65....1K,2010ASPC..421...11K}. Isolated galaxies represent the nurture-free zero point for characterising secular processes \citep{2007A&A...470..505V,2007A&A...472..121V}. The observed properties of isolated galaxies are likely to be determined mainly by their initial formation conditions and secular evolutionary processes \citep{1981Afz....17...53A,1987MNRAS.226..543E,1988ApJ...334..613S,2011RMxAA..47..361C,2012A&A...545A..15S,2014ApJ...788L..39F}. 

Separating the effects of interaction-driven evolution in the observed galaxy properties is not trivial. Galaxies suffer intrinsic and secular evolution processes, but they are also exposed to the influences of their local and large-scale environments. There is a strong correlation of galaxy properties with the environment, such as the morphology-density relation \citep{1980ApJ...236..351D}; optical and ultraviolet luminosities and atomic gas mass function \citep{2009ARA&A..47..159B,2011MNRAS.418...64D}; the morphology-mass relation \citep{2011MNRAS.tmpL.354C}; the infrared luminosity and stellar-mass function \citep{2001ApJ...557..117B}; galaxy colours \citep[][and references therein]{2011MNRAS.411..929G}; the rate of accretion events \citep{2010MNRAS.407.1514E,2012AJ....144..128A}; and  nuclear activity \citep{2004MNRAS.353..713K,2009ApJ...699.1679C,2013MNRAS.430..638S}. Additional studies are needed to separate the effect of one-on-one interactions from the influence of the LSS. Therefore, the study of field systems, such as galaxy pairs and galaxy triplets, is also helpful for understanding the galaxy evolution on intermediate scales between isolated galaxies and denser environments (i.e., groups and clusters). 

The new 3D surveys are required to disentangle small, faint, linked satellites from a background projected galaxy population, which leads to a more comprehensive 3D picture of the surroundings. The construction of two-dimensional (2D) photometric-only catalogues of galaxies presents a large uncertainty introduced by projection effects. For instance, in a previous work \citep{2014A&A...564A..94A}, we describe the 3D environment for a sample of 386 galaxies in the CIG using the Ninth Data Release of the SDSS \citep[DR9;][]{2012ApJS..203...21A}. We found that almost all GIG galaxies (97\%) are distributed following the LSS of the local Universe with a large heterogeneity in their degrees of connection with it, from galaxies in interaction with physical satellites (15\%) to galaxies with no neighbours within 3\,Mpc (less than 3\%). For this reason, we aim to compile and characterise samples of isolated galaxies, isolated pairs, and isolated triplets in the local Universe without being biased by projected neighbours, by applying the lessons we have learned in previous works. These catalogues will represent valuable samples for future studies of the dependence of different galaxy properties on both the local and large-scale environments.

This study is organised as follows. In Sect.~\ref{Sec:data3} the data and the isolation criteria used to compile the different samples are presented. The methods of identifying the physically bound pairs and triplets are described in Sect.~\ref{Sec:physical3}. In Sect.~\ref{Sec:quantification3} we introduce the parameters used to quantify the environment of the galaxies in the catalogues. We present our results in Sect.~\ref{Sec:results3} and the associated discussion in Sect.~\ref{Sec:discussion3}. Finally a summary and the main conclusions of the study are presented in Sect.~\ref{Sec:con3}. Throughout the study, a cosmology with $\Omega_{\Lambda 0} = 0.7$, $\Omega_{\rm{m} 0} = 0.3$, and $H_{0}=70$\,km\,s$^{-1}$\,Mpc$^{-1}$ is assumed.

\section{Data and definition of the isolation}         \label{Sec:data3}

The data is based on the Tenth Data Release of the SDSS \citep[DR10;][]{2013arXiv1307.7735A}. We defined a primary sample composed of galaxies from the main spectroscopic sample  \citep{2002AJ....124.1810S} with $r$-band (the deepest SDSS images) model magnitudes in the range $11 \leq m_{r} \leq 15.7$ (sufficient to develop a homogeneous isolation definition fainter within at least 2 magnitudes than the primary, since the redshift completeness of the SDSS is $m_{r,\rm{Petrosian}}~<~17.77$\,mag), and over a redshift range $0.005 \leq z \leq 0.080$. We discarded galaxies with redshift lower than 0.005 to avoid nearby sources for which isolation cannot be estimated well \citep{2007A&A...470..505V} and larger errors from photometry for very extended galaxies. Galaxies with a redshift above 0.080 were also discarded to facilitate future studies based on visual morphological classifications.

After compiling a first sample of primary galaxies using these conditions, we proceeded to clean the sample following a pipeline similar to the one developed in \citet{2014A&A...564A..94A}: (a) we selected all galaxies within at least 1\,Mpc radius completely sampled by the SDSS-DR10 photometric footprint, (b) we performed a star-galaxy separation in order to reject stars misclassified as galaxies in the automated SDSS pipelines, and (c) we applied an automated cleaning process to discard multiple identifications of the same galaxy.

To characterise the environment of primary galaxies, we used the CasJobs\footnote{\texttt{http://skyservice.pha.jhu.edu/CasJobs/}} tool to search for neighbours (hereafter the neighbour sample) within 1\,Mpc around each candidate isolated galaxy in the primary sample and with redshift between 0.001 and 0.100 \citep[in either the SDSS main galaxy spectroscopic sample or the Baryon Oscillation Spectroscopic Survey; BOSS,][]{2013AJ....145...10D}. Star-galaxy separation and cleaning for multiple identifications as in \citet{2014A&A...564A..94A} were also performed to avoid duplication in the neighbour sample. 

The SDSS spectroscopic main galaxy sample is supposed to be complete in the range $14.5~\lesssim~m_{r}~\lesssim~17.7$ \citep{2002AJ....124.1810S}. We minimised the effects of redshift incompleteness by selecting primary galaxy fields where the percentage of extended neighbours within 1\,Mpc with measured spectroscopic redshift is higher than 80\%. On this basis we discarded 24\% of the fields.

The final primary galaxy sample was composed of 33~081 galaxies surrounded by 1~607~947 neighbours. These compiled samples were used to extract catalogues of isolated galaxies, isolated pairs, and isolated triplets. We searched for galaxies in the primary sample with, respectively, zero, one, and two companions in the neighbour sample and within radial velocity differences $-500~\leq~\Delta \varv~\leq~500$\,km\,s$^{-1}$ (and within 1\,Mpc) from the primary galaxy. Our isolation criterion is therefore based on selection in a velocity difference--projected distance space.

We removed multiple identifications for the same pair and/or triplet candidates by choosing unique systems where the brightest galaxy is in the primary sample. This brightest galaxy must have one or two neighbours according to the isolation criteria, if the system is a candidate for isolated pair or isolated triplet, respectively. Using these criteria, we found a total of 3~867 isolated galaxy, 4~002 isolated pair, and 3~052 isolated triplet candidates.

Not all galaxies in the magnitude range  $14.5~\lesssim~m_{r}~\lesssim~17.7$ are in the SDSS spectroscopic sample. Owing to fibre collisions \citep[two fibres of the SDSS spectrograph cannot be placed closer than 55\arcsec,][]{2002AJ....124.1810S} close galaxy pairs or mergers cannot always be independently identified as distinct targets. To amend this, we therefore visually revised the SDSS three-colour images of each candidate in the previous samples. We also used the environmental search of the NASA/IPAC Extragalactic Database (NED\footnote{\texttt{http://ned.ipac.caltech.edu/}}) to search for the redshift of possible companions without spectroscopic information in the SDSS. On this basis we removed 201 isolated galaxy candidates, 155 isolated pair candidates, and 94 isolated triplet candidates.

Out of the 201 isolated galaxy candidates removed, 53 were candidate mergers. Out of the 155 pair candidates removed, we identified 35 as candidate isolated galaxy since the``'neighbour'' is actually part of the same primary galaxy (i.e., an H$_{\rm{II}}$ region or a source missed by our cleaning pipeline in very extended primaries). Similarly, out of the 94 triplet candidates removed, one was identified as an isolated galaxy (the two other components of the ``triplet'' are found to be H$_{\rm{II}}$ regions), and 29 were identified as pairs. The final numbers of isolated system candidates are 3~702 isolated galaxies, 3~876 isolated pairs, and 2~958 isolated triplets.

\section{Physical definition}  \label{Sec:physical3}

Isolated galaxies are assumed, by definition, to have escaped serious influences from  their nearest neighbours during a crossing time $t_{\rm{cc}} \approx $\,3\,Gyr \citep{2005A&A...436..443V}. When assuming a typical "field" velocity dispersion of the order of 190\,km\,s$^{-1}$ \citep{2000ApJ...530..625T}, a crossing time of about $t_{\rm{cc}}\sim\,5$\,Gyr is implied for crossing the selected field radius of 1\,Mpc. Isolation degree as defined in this study is more conservative than the previous definition connected with the CIG sample. The adopted catalogue of isolated galaxies is composed of the 3~702 entries considered to have been isolated a majority of their lifetimes.

\begin{figure}
\centering
\includegraphics[width=\columnwidth]{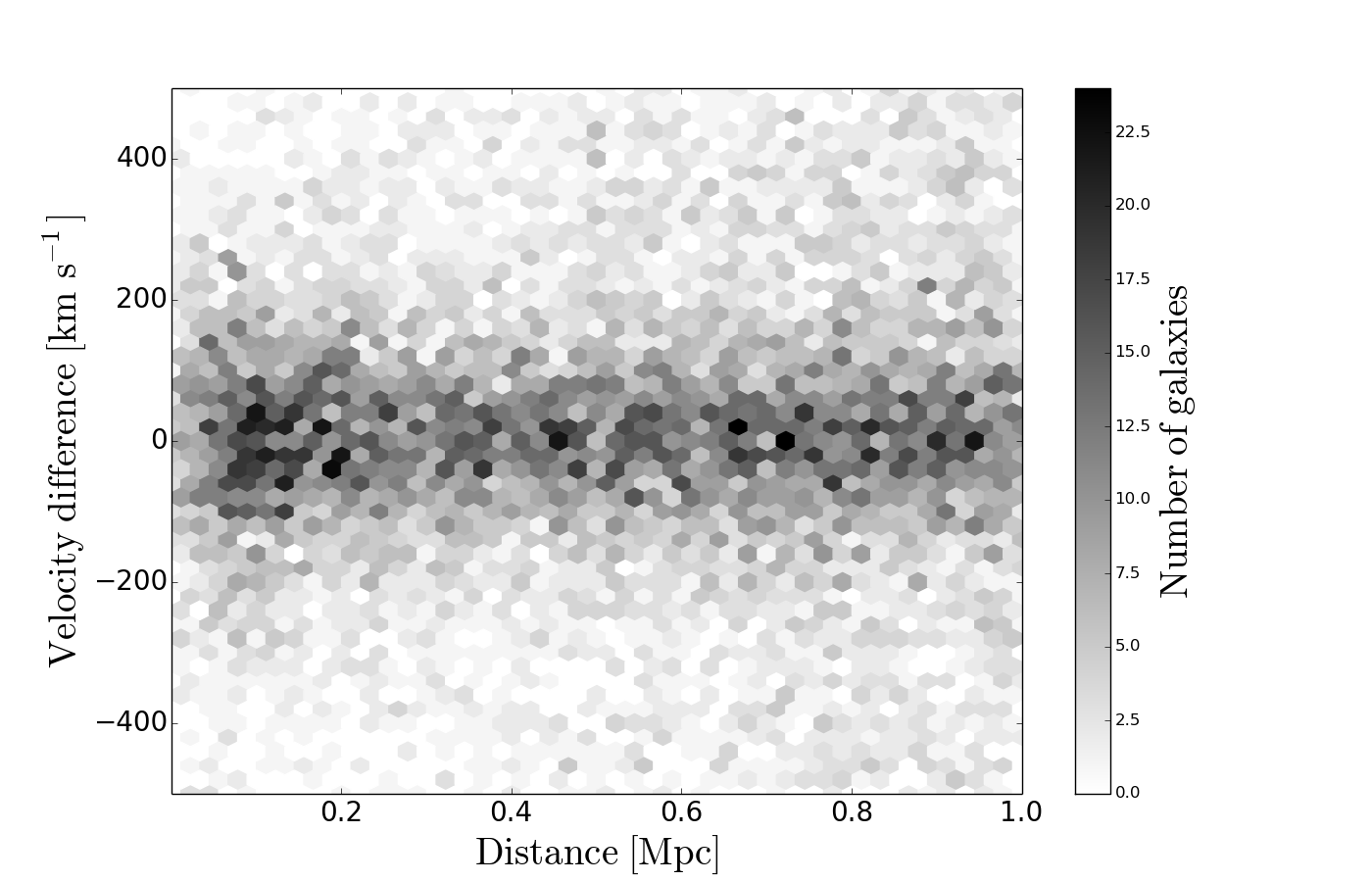} \\
\includegraphics[width=\columnwidth]{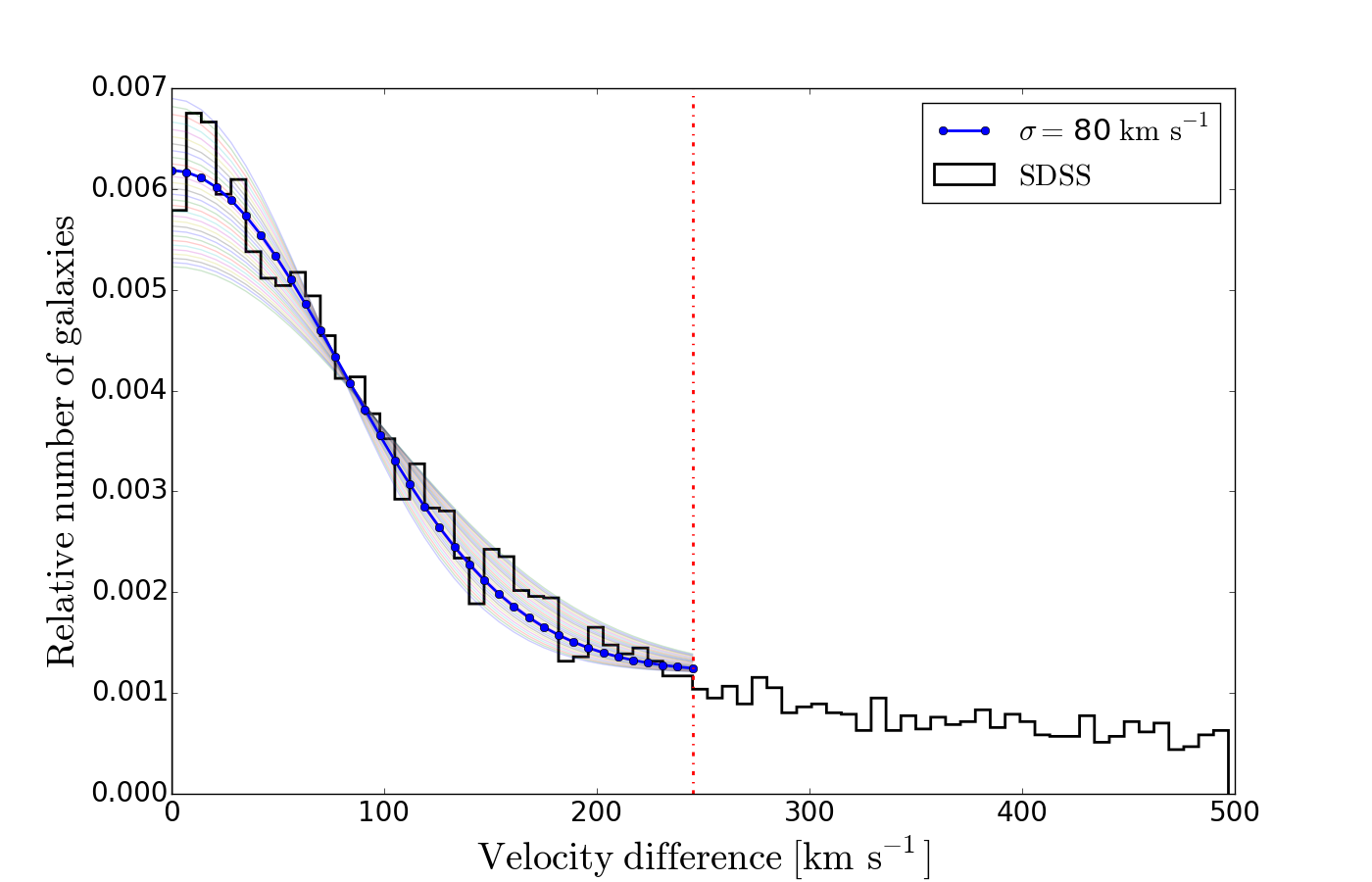} \\
\includegraphics[width=\columnwidth]{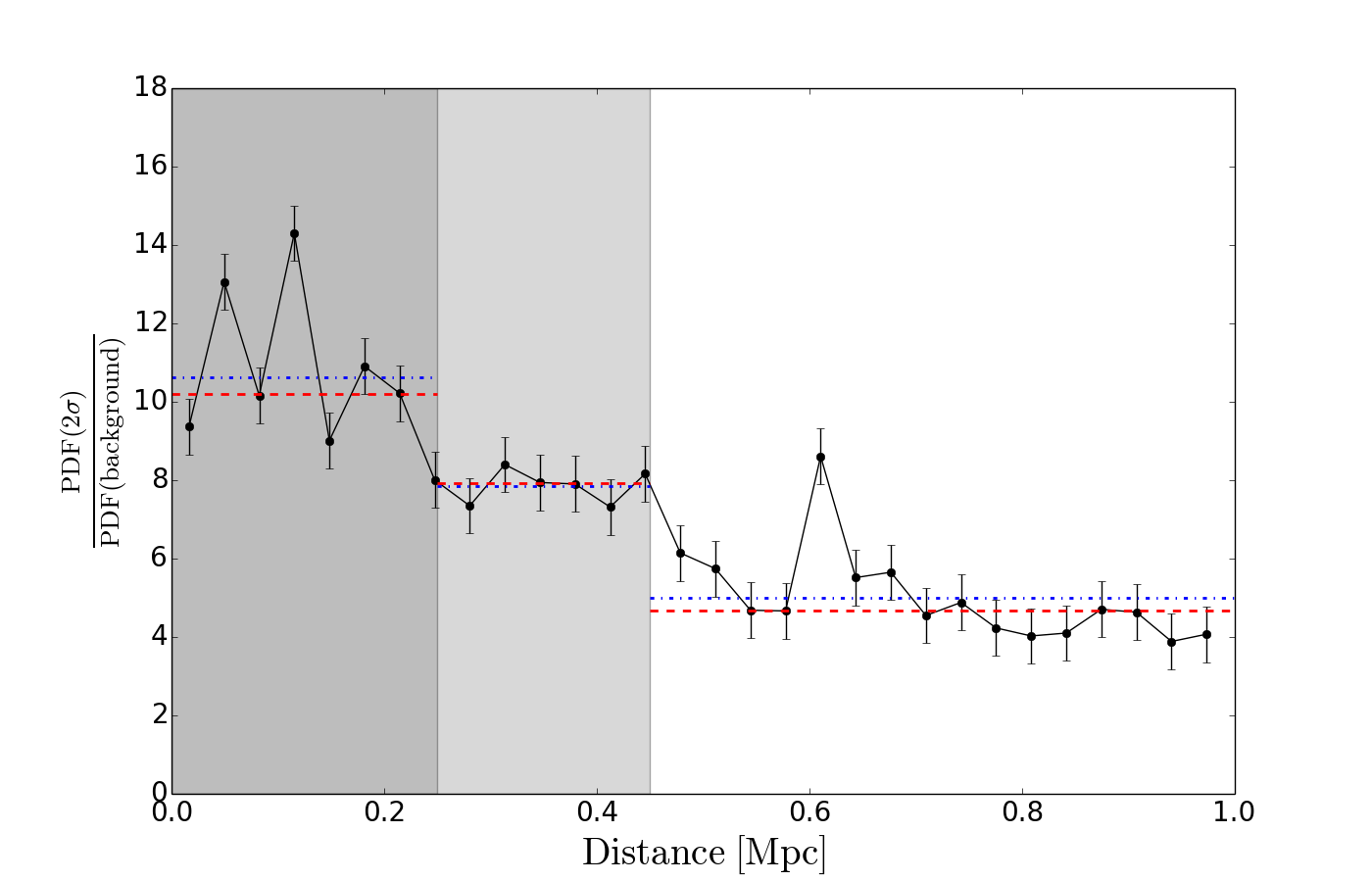}
\caption[Selection of physical pairs and triplets]{Selection of physical pairs and triplets. {\it (Upper panel):} 2D distribution of the line-of-sight velocity difference and projected distance for isolated-pair and isolated-triplet candidates with respect to the brightest \emph{A} galaxy. Colours correspond to the galaxy counts according to the colour bar. {\it (Middle panel):} Absolute values of the line-of-sight velocity difference $\Delta\,\varv = |\varv_{\rm{A}} - \varv_{N}|$ (black histogram), and the corresponding Gaussian distribution fits for $\sigma$ between 70 and 100\,km\,s$^{-1}$ (grey curves) with the best fit (blue curve) within the first 250\,km\,s$^{-1}$ (red dashed line). The corresponding $2\sigma$ is shaded. {\it (Lower panel):} Probability density function (PDF) for neighbour galaxies within $2\sigma$ over PDF for the background flat population selected in the interval $300 < \Delta\,\varv < 500$\,km\,s$^{-1}$, as a function of the distance to the \emph{A} galaxy. Red dashed and blue dash-dotted lines correspond to the median and mean values in the different distance intervals, respectively. The inner regions up to 250\,kpc and 450\,kpc are shaded.}
\label{Fig:physicals3}
\end{figure}

The 2D distribution of $\Delta\,\varv$ and distance $d$ (from the brightest galaxy, hereafter \emph{A} galaxy) for isolated pair and triplet candidates, shows an over-density, approximately at $-200~\lesssim~\Delta\,\varv~\lesssim~200$\,km\,s$^{-1}$ and $d~\lesssim~500$\,kpc (see the upper panel of Fig.~\ref{Fig:physicals3}). This means that neighbour galaxies within these limits are likely to be physically connected to their corresponding \emph{A} galaxy. Conversely, neighbour galaxies showing higher $\Delta\,\varv$ and $d$ would very likely be related to the underlying large-scale distribution of galaxies. 

As the previous work of \citet{2014A&A...564A..94A}, the distribution of line-of-sight velocity differences with respect to the \emph{A} galaxy, in this case for isolated pairs and isolated triplets candidates, shows a Gaussian distribution. We follow a similar procedure to identify the physically bound companions by analysing the location of the neighbours within the distribution with respect to the \emph{A} galaxy. The middle panel of Fig.~\ref{Fig:physicals3} shows an estimation of the standard deviation $\sigma$ of the $\Delta\,\varv~\equiv~|\varv_{A} - \varv_{N}|$ distribution, where $\varv_{A}$ and $\varv_{N}$ are the line-of-sight velocities of the \emph{A} galaxy and neighbour $N$, respectively. The best Gaussian distribution fit gives a value of $\sigma = 80$\,km\,s$^{-1}$. The neighbour galaxies within $\Delta\,\varv \leq 2 \sigma$ (95.4\% of the physically linked companions) show a tendency to be located within the first 450\,kpc from the \emph{A} galaxy (see the lower panel of Fig.~\ref{Fig:physicals3}). Inside this area, another over-density region can be detected from 0 to 250\,kpc (shaded inner region in lower panel of Fig.~\ref{Fig:physicals3}). We consider that the non-zero excess of similar redshift galaxies at larger distances is due to galaxies of their associated LSS. 

To select physically bound isolated systems, we considered isolated pairs where the faintest galaxy, hereafter \emph{B}, has $\Delta \varv \leq 160$\,km\,s$^{-1}$ ($2 \sigma$) and $d \leq 450$\,kpc (shaded outer region in lower panel of Fig.~\ref{Fig:physicals3}). In the case of isolated triplets, we required that the two fainter galaxies, hereafter \emph{B} and \emph{C} galaxies, both satisfy $\Delta \varv \leq 160$\,km\,s$^{-1}$ and $d \leq 450$\,kpc.

\section{Quantification of the environment}  \label{Sec:quantification3}

We used two parameters to quantify the effects of one-to-one interactions and the large-scale environment on galaxy properties \citep{2007A&A...472..121V,2013MNRAS.430..638S,2013A&A...560A...9A,2014A&A...564A..94A}: the tidal strength parameter $Q$ (Eq.~\ref{Eq:Q3}) and the projected density $\eta_{k}$ (Eq.~\ref{Eq:etak3}).

\subsection{Tidal strength}

The tidal strength parameter $Q$ is an estimation of the total gravitational interaction strength that the neighbours produce on the central galaxy with respect to its internal binding forces. The tidal strength on a primary galaxy $P$ created by all the neighbours $i$ in the field is

\begin{equation} \label{Eq:Q3}
Q \equiv {\rm log} \left(\sum_i {\frac{M_{i}}{M_{P}}} \left(\frac{D_P}{d_i}\right)^3\right) \quad,
\end{equation} 
where $M$ is the stellar mass of the galaxy and $d_i$  the projected physical distance of the $i^{\rm th}$ neighbour to the primary galaxy. Stellar masses were calculated by fitting the spectral energy distribution on the five SDSS bands using the routine kcorrect \citep{2007AJ....133..734B}. The parameter $D_P = 2\alpha r_{90}$ is the estimated diameter of the primary galaxy, where $r_{90}$, the Petrosian radius containing 90\,\% of the total flux of the galaxy in the $r$ band, is scaled by a factor $\alpha=1.43$ to recover the $D_{25}$ \citep{2013A&A...560A...9A}. The greater the value of $Q$, the less isolated from external influence the primary galaxy. The value of $Q$ is flagged as ``NULL'' when there is no neighbour to estimate it.

\subsection{Projected density}

To characterise the LSS around the primary galaxies in the catalogues, we also defined the projected number density parameter as
\begin{equation} \label{Eq:etak3}
\eta_{k, \rm LSS} \equiv \log \left(\frac{k - 1}{\rm{Vol}(d_k)}\right) = \log \left(\frac{3(k - 1)}{4\pi\,d_k^3}\right)\quad,
\end{equation} 
where $d_k$ is the projected physical distance to the $k^{\rm th}$ nearest neighbour, with $k$ equal to 5, or less if there were not enough neighbours in the field. The farther the $k^{\rm th}$ nearest neighbour, the lower the projected number density $\eta_{k, \rm LSS}$. The value of $\eta_{k, \rm LSS}$ is flagged as ``NULL'' when there are one or zero neighbours. For isolated pairs and isolated triplets, the components of the system are excluded for estimating the projected density, so we consider these systems as a whole when studying their relation to the LSS.


\section{Results}\label{Sec:results3}

\subsection{The catalogues} \label{Sec:res_cat3} 

We compiled catalogues of isolated galaxies, hereafter SIG\footnote{SDSS-based Isolated Galaxies}, isolated pairs, hereafter SIP\footnote{SDSS-based Isolated Pairs}, and isolated triplets, hereafter SIT\footnote{SDSS-based Isolated Triplets} in the local Universe. We identified 3~702 isolated galaxies (see Table~\ref{tab:tab1}) in the primary catalogue (see Sect.~\ref{Sec:data3}) with no neighbours within $\Delta\,\varv\,\leq\,500$\,km\,s$^{-1}$ and $d~\leq~1$\,Mpc. The SIG sample represents about 11\% of the total number of galaxies in the local Universe ($z\leq0.080$) considered in the study.

For the final pair and triplet catalogues, we followed a similar criteria to \citet{2014A&A...564A..94A} and considered that the over-density of similar redshift galaxies found in the lower panel of Fig.~\ref{Fig:physicals3} is due to the nature of the physically bound systems. Therefore the bound isolated systems are selected when $\Delta\,\varv~\leq~160$\,km\,s$^{-1}$ ($2 \sigma$) and $d~\leq~450$\,kpc. 

These selection criteria lead to 1~240 isolated pairs (see Table~\ref{tab:tab2}) and 315 isolated triplets (see Table~\ref{tab:tab3}). The SIP and SIT samples represent about 7\% and 3\% of the galaxies in the local Universe considered in the study, respectively. 
Galaxies in pairs and triplets represent almost half of the galaxies considered as field galaxies in the local Universe (26\%). Figures~\ref{Fig:charts10isol}, \ref{Fig:charts10pairs}, and \ref{Fig:charts10triplets} show some SDSS images of galaxies in the catalogue of isolated galaxies, isolated pairs, and isolated triplets, respectively.

\begin{figure}
\centering
\includegraphics[width=\columnwidth]{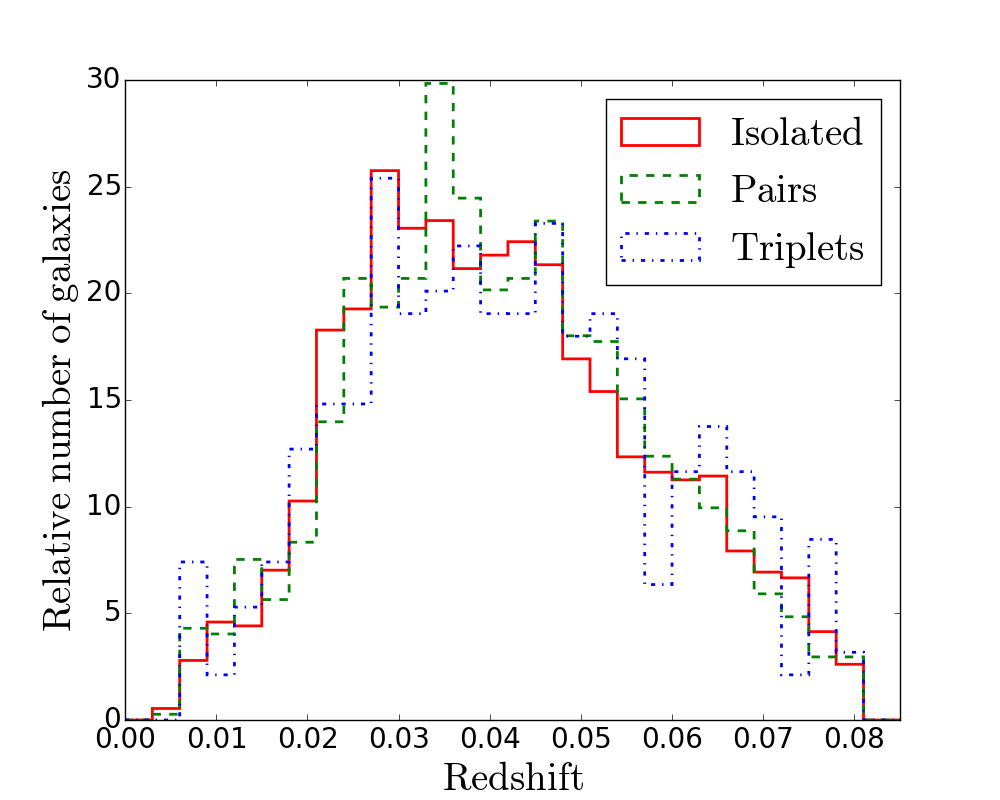}
\caption[Redshift distribution for isolated, pair, and triplet galaxies]{Redshift distribution for isolated galaxies (red solid histogram), isolated pairs (green dashed histogram), and isolated triplets (blue dash-dotted histogram).}
\label{Fig:histz3}
\end{figure}

The distribution of the redshifts for isolated galaxies, isolated pairs, and isolated triplets (see Fig.~\ref{Fig:histz3}) shows that there is no preferred location (there is no trend with redshift for the three samples), which suggests that there is no obvious concentration associated with major components of large-scale structure. To visualise this resultbetter  we have developed the interactive visualisation tools described in Appendix~\ref{Sec:App1}.

\begin{table*}[!htbp]
  \caption{\label{tab:tab1}SDSS-based catalogue of 3~702 isolated galaxies (SIG) in 3D and relation with their large-scale environment.} 
\centering
\begin{tabular}{ccccccccc}
\hline \hline
(1) & (2) & (3) & (4) &  (5) & (6) & (7) & (8) & (9) \\
 SIG & RA & DEC & $z$ &  $k_{\rm{LSS}}$  &  $f_{\rm{LSS}}$  &  $d_{\rm{NN}}$  &   $\eta_{k,\rm{LSS}}$  &  $Q_{\rm{LSS}}$ \\
    & (deg) & (deg) &  &  & &  (Mpc) &  &  \\
\hline
     1     &    148.87999     &     -0.46724     &      0.03221     &    36     &     1     &     1.50     &    -0.77     &    -5.78    \\
     2     &    148.18303     &     -0.05831     &      0.04565     &    20     &     1     &     2.39     &    -1.55     &    -5.42    \\
     3     &    149.61346     &      0.38248     &      0.05648     &     3     &     1     &     3.32     &    -1.98     &    -6.15    \\
     4     &    147.45673     &      1.09258     &      0.03480     &    23     &     1     &     2.57     &    -1.49     &    -5.18    \\
     5     &    147.32950     &      0.02890     &      0.04809     &    12     &     1     &     1.85     &    -1.89     &    -5.70    \\
     6     &    188.06413     &     65.76231     &      0.04677     &    11     &     1     &     1.77     &    -1.04     &    -5.50    \\
     7     &    190.39415     &     65.71832     &      0.03857     &     1     &     1     &     2.72     &     NULL     &    -7.18    \\
     8     &    126.44534     &     49.57347     &      0.02376     &    12     &     1     &     1.24     &    -0.78     &    -4.71    \\
     9     &    132.28497     &     53.78013     &      0.02453     &    19     &     1     &     1.25     &    -0.83     &    -5.36    \\
    10     &    258.53766     &     58.81841     &      0.02978     &    75     &     1     &     1.08     &    -0.75     &    -4.09    \\
\ldots & \ldots & \ldots & \ldots & \ldots  & \ldots & \ldots & \ldots  & \ldots \\
\hline   
\end{tabular}
\tablefoot{The full table is available in electronic form at the CDS. 
The columns correspond to: 
(1) isolated galaxy identification; 
(2) J2000.0 right ascension in degrees; 
(3) J2000.0 declination in degrees; 
(4) redshift of the galaxy; 
(5) $k_{\rm{LSS}}$, number of LSS associations; 
(6) $f_{\rm{LSS}}$, footprint flag: ``1'' if the galaxy has a 5\,Mpc field radius in the SDSS footprint, ``0'' if not;
(7) $d_{\rm{NN}}$, distance to the first nearest neighbour from 1\,Mpc to 5\,Mpc; 
(8) $\eta_{k,\rm{LSS}}$, projected density estimation of the LSS; 
(9) $Q_{\rm{LSS}}$, tidal strength estimation of the LSS.}
\end{table*}


\subsection{Comparison with previously existing catalogues} \label{Sec:res_others}

There are other extant catalogues of isolated galaxies, isolated pairs, and isolated triplets available in the literature. Unfortunately, a direct comparison of the properties of the galaxies with them becomes very complicated owing to the different nature of the samples (selection criteria, database, wavelength, etc.). The homogeneity in the selection criteria of the SIG, SIP, and SIT catalogues permits an easier comparison and interpretation of the results. Nevertheless, as a first step, we need to understand the similarities and differences of our catalogues with respect to the 2D and 3D catalogues in the literature. For this comparison, we use publicly available catalogues based on visual 2D isolation criteria and catalogues based on SDSS 3D information for consistency with our work.

Pioneering works that identify isolated galaxy systems in 2D, before the era of large data sets from redshift galaxy surveys, were performed by \citet{1973SoSAO...8....3K} for a catalogue of isolated galaxies (hereafter CIG), \citet{1972SoSAO...7....1K} for a catalogue of isolated pairs (hereafter KPG), and \citet{1979AISAO..11....3K} for a catalogue of isolated triplets (hereafter KTG). It is worthwhile mentioning that the CIG, KPG, and KTG were compiled using a 2D visual isolation criterion based on comparing galaxy apparent diameters with the foreground and background galaxies. 

Recently, other catalogues of isolated galaxies, isolated pairs, and isolated triplets have been compiled in 3D using digital data and are publicly available \citep{2006AJ....132.2243G,2009AstBu..64...24M,2010Ap.....53..462K,2010AJ....139.1857F}. For consistency, we selected samples based on SDSS spectroscopy to compare with our catalogues of isolated galaxies, isolated pairs, and isolated triplets. We selected the UNAM-KIAS catalogue of isolated galaxies \citep{2010AJ....139.2525H} based on SDSS-DR5 data, where the authors applied a CIG-like isolation criterion with the addition of a third condition about velocity difference between central galaxy and neighbours. To compare with our catalogue of isolated pairs, we chose the sample of pairs (with projected separations $d<5-20~h^{-1}$\,kpc) compiled by \citet{2008ApJ...685..235P} to study the luminosity dependence on pair distance which is based on data from SDSS-DR5. We used the catalogue of triplets compiled by \citet{2012MNRAS.421.1897O}, based on SDSS-DR7 spectroscopic data, to compare with our catalogue of isolated triplets. 

We based our comparison on the galaxies in common with the primary sample (sample of candidates before applying the isolation criteria), i.e., how many galaxies satisfy the same selection criteria in magnitude and redshift ranges, field coverage, and redshift completeness, as explained in Sect.~\ref{Sec:data3}. Then we checked how many of the primary galaxies are found to be the \emph{A} galaxy in the SIG, SIP, and SIT catalogues. The results of the comparisons are listed in Table~\ref{tab:comparison}. The CIG is originally composed of 1050 galaxies, and we find 275 CIG galaxies in the primary sample. Out of them we find 79 in our catalogue of isolated galaxies. In comparison with the KPG, composed of 603 galaxy pairs, we find 170 in our primary sample, and only three of them are in our isolated pairs. In the case of galaxy triplets, the KTG is composed of 84 triplets, out of which we find 20 in the primary sample, and only one KTG in our 3D isolated triplets. Similarly, in comparison with the catalogues based on SDSS data, there are 449 isolated galaxies, three isolated pairs, and one isolated triplet in common (see discussion in Sect.~\ref{Sec:dis_others}).

\begin{table}
  \caption{\label{tab:tab2}SDSS-based catalogue of 1~240 isolated pairs (SIP) in 3D.} 
\centering
\begin{tabular}{ccccc}
\hline \hline
(1) & (2) & (3) & (4) &  (5)  \\
 SIP & index & RA & DEC & $z$ \\
     &    & (deg) & (deg) &  \\
\hline
     1     &     1     &    146.29984     &     -0.12000     &      0.03070   \\
     1     &     2     &    146.35577     &     -0.14337     &      0.03074   \\
     2     &     1     &    148.97949     &     -0.19028     &      0.02095   \\
     2     &     2     &    149.09021     &     -0.25143     &      0.02140   \\
     3     &     1     &    130.45488     &     50.78642     &      0.05439   \\
     3     &     2     &    130.41447     &     50.78843     &      0.05434   \\
     4     &     1     &    152.77184     &      1.22408     &      0.03315   \\
     4     &     2     &    152.77502     &      1.22543     &      0.03308   \\
     5     &     1     &    150.46953     &      3.01132     &      0.04409   \\
     5     &     2     &    150.42798     &      3.00499     &      0.04407   \\
\ldots & \ldots & \ldots & \ldots & \ldots \\
\hline   
\end{tabular}
\tablefoot{The full table is available in electronic form at the CDS. 
The columns correspond to: 
(1) isolated pair identification; 
(2) index of the galaxy in the pair: ``1'' for the \emph{A} galaxy, ``2'' for the \emph{B} galaxy; 
(3) J2000.0 right ascension in degrees; 
(4) J2000.0 declination in degrees; 
(5) redshift of the galaxy.}
\end{table}

\begin{table}
  \caption{\label{tab:tab3}SDSS-based catalogue of 315 isolated triplets (SIT) in 3D.} 
\centering
\begin{tabular}{ccccc}
\hline \hline
(1) & (2) & (3) & (4) &  (5) \\
 SIT & index & RA & DEC & $z$  \\
    &    & (deg) & (deg) &  \\
\hline
     1     &     1     &    171.31543     &     -2.07612     &      0.06287     \\  
     1     &     2     &    171.40068     &     -2.06681     &      0.06300     \\  
     1     &     3     &    171.26364     &     -2.09105     &      0.06297     \\  
     2     &     1     &    208.67705     &     65.24431     &      0.03578     \\  
     2     &     2     &    209.01042     &     65.29638     &      0.03615     \\  
     2     &     3     &    208.66452     &     65.27580     &      0.03565     \\  
     3     &     1     &    211.20346     &      4.98106     &      0.04649     \\  
     3     &     2     &    211.19240     &      4.97013     &      0.04695     \\  
     3     &     3     &    211.28189     &      4.96810     &      0.04647     \\  
\ldots & \ldots & \ldots & \ldots & \ldots  \\
\hline   
\end{tabular}
\tablefoot{The full table is available in electronic form at the CDS. 
The columns correspond to: 
(1) isolated triplet identification; 
(2) index of the galaxy in the triplet: ``1'' for the \emph{A} galaxy, ``2'' for the \emph{B} galaxy, and ``3'' for the \emph{C} galaxy; 
(3) J2000.0 right ascension in degrees; 
(4) J2000.0 declination in degrees; 
(5) redshift of the galaxy.}
\end{table}

\begin{table}
  \caption[Comparison samples]{\label{tab:comparison}Comparison samples.} 
\centering
\begin{tabular}{lcccc}
\hline \hline
 Sample & Initial & Primary & Final & Percentage\\
\hline
CIG (isolated)       &    1050     &     275     &      79     &    29\%    \\
KPG (pairs)          &     603     &     170     &       3     &    2\%     \\
KTG (triplets)       &      84     &      20     &       1     &    5\%     \\
\hline
UNAM-KIAS (isolated) &    1520     &    1031     &      449    &    43.5\%  \\
Patton+08 (pairs)    &     473     &      41     &       3     &    7\%     \\
O'Mill+12 (triplets) &    1092     &      20     &       1     &    5\%     \\
\hline   
\end{tabular}
\tablefoot{The line separates samples that use the SDSS (lower part) from samples that do not (upper part). The columns correspond to: 
(1) name of the catalogue; 
(2) number of galaxy systems in the original catalogue; 
(3) number of (central) galaxies in common with our primary sample (sample of candidates before applying the isolation criteria); 
(4) number of (central) galaxies found in the SIG, SIP, or SIT catalogues;
(5) percentage with respect to the galaxies in common to the fourth column.}
\end{table}


\subsection{Relation to the large-scale structures} \label{Sec:res_LSS3}

To understand the effects of the large scale environment on galaxy properties, we characterise the LSS around each primary galaxy up to 5\,Mpc. We consider part of the LSS the neighbour galaxies within $\Delta \varv \leq 500$\,km\,s$^{-1}$ from 1 to 5\,Mpc. 

Not all the galaxies in the catalogues have a 5\,Mpc field radius that is completely covered by the SDSS footprint. To identify these galaxies, we added the column $f_{\rm{LSS}}$ with value ``1'' if the galaxy has a 5\,Mpc field radius within the SDSS footprint, and ``0'' if not, in Tables~\ref{tab:tab1},~\ref{tab:tab2_2}, and~\ref{tab:tab3_2}. 

There are 3~535 isolated galaxies, 1~182 isolated pairs, and 301 isolated triplets within the SDSS footprint at 5\,Mpc, which corresponds to 95.5\% of the galaxies in each sample. Out of them, we find 31 isolated galaxies, 9 isolated pairs, and 0 isolated triplet with no neighbours within $\Delta \varv \leq 500$\,km\,s$^{-1}$ up to 5\,Mpc. We check that these galaxies are not mainly located in voids (we discuss this further in Sect.~\ref{Sec:dis_LSS}) also using the visualisation tools described in Appendix~\ref{Sec:App1}.

\begin{figure*}
\begin{center}
\includegraphics[width=.49\textwidth]{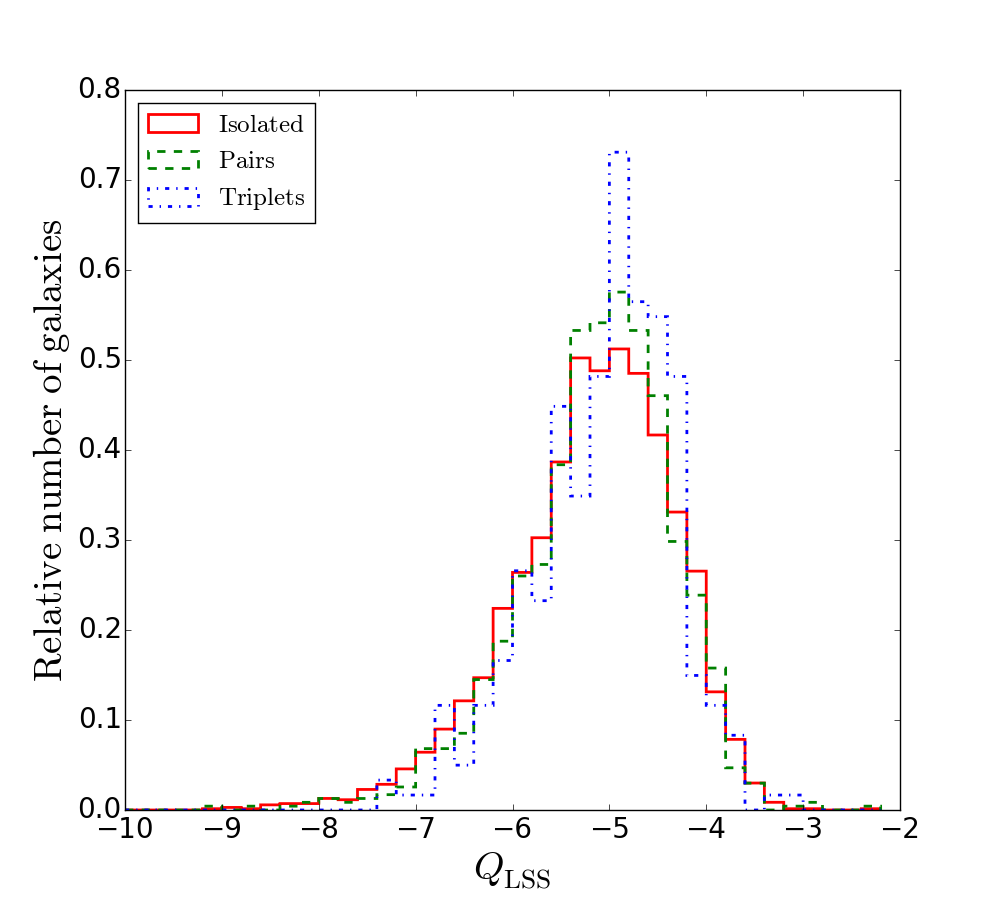} 
\includegraphics[width=.49\textwidth]{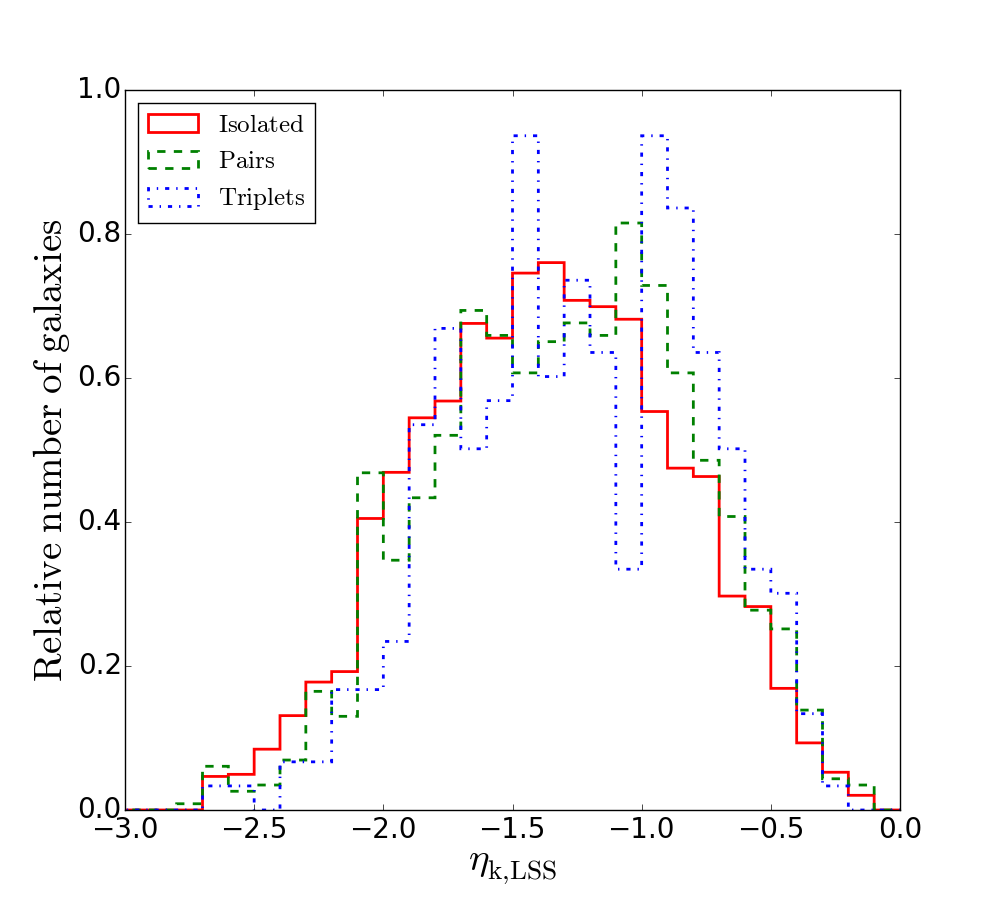} \\
\includegraphics[width=.49\textwidth]{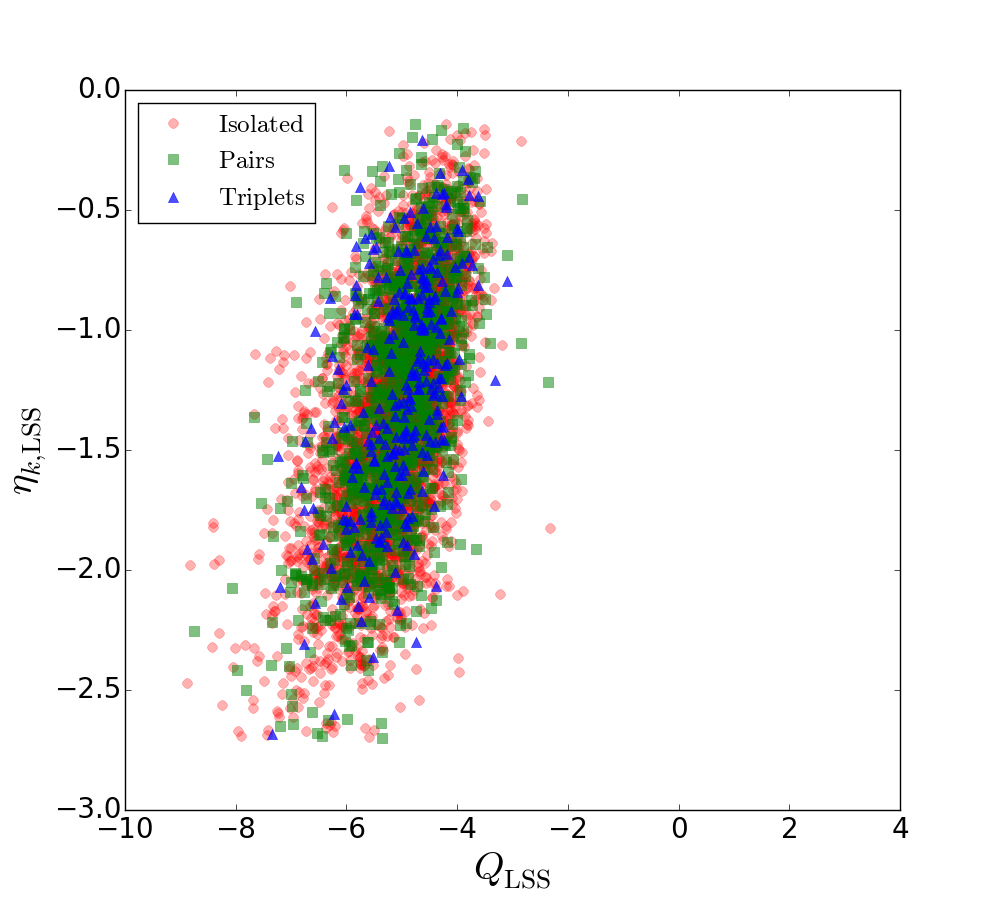} 
\includegraphics[width=.49\textwidth]{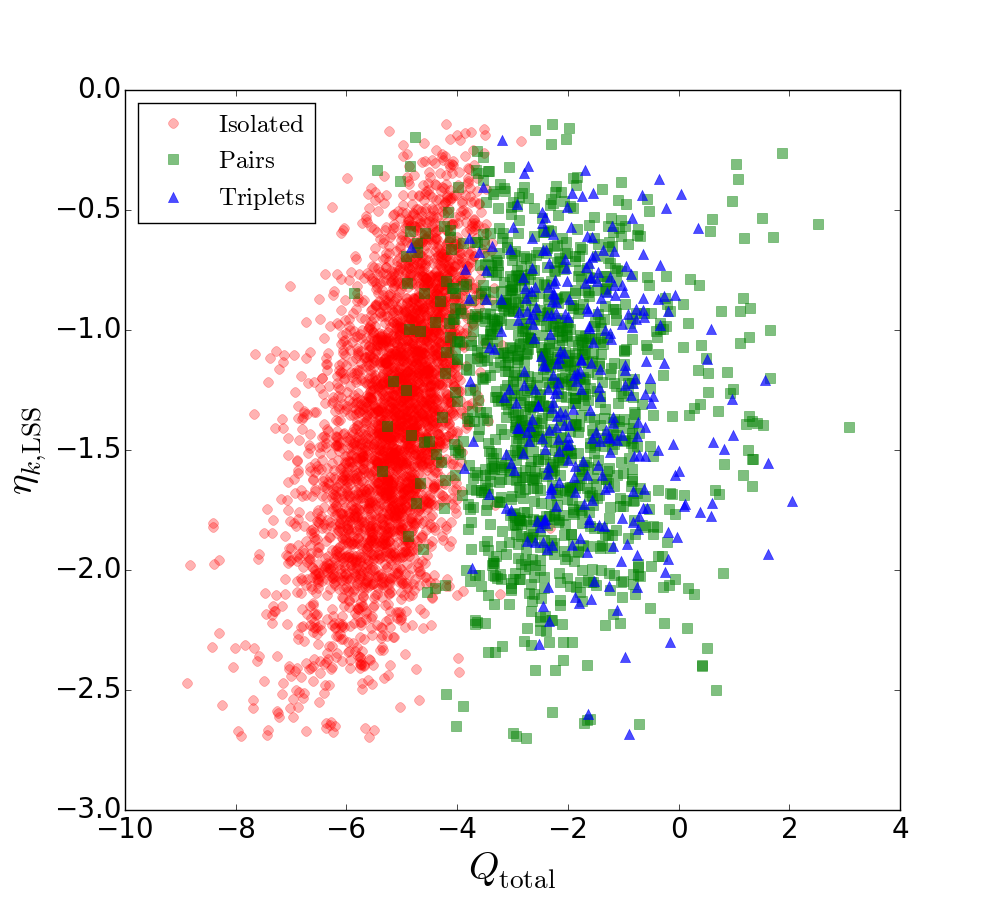} 
\end{center}
\caption{LSS isolation parameters for the SIG, SIP, and SIT catalogues. {\it (Upper left panel):} Distribution of the tidal strength $Q_{\rm LSS}$ for isolated galaxies (red solid histogram), isolated pairs (green dashed histogram), and isolated triplets (blue dash-dotted histogram). {\it (upper right panel):} Distribution of the projected density $\eta_{k,\rm LSS}$ in the same colour scheme. {\it (Lower left panel):} Projected density $\eta_{k, \rm LSS}$ versus tidal strength $Q_{\rm LSS}$ diagram. Red circles, green squares, and blue triangles represent isolated galaxies, isolated pairs, and isolated triplets, respectively. {\it (Lower right panel):} Projected density $\eta_{k, \rm LSS}$ versus tidal strength $Q_{\rm total}$ diagram. Red circles, green squares, and blue triangles represent isolated galaxies, isolated pairs, and isolated triplets, respectively.} \label{Fig:isolparam3}
\end{figure*}

The tidal strength generated by the galaxies in the LSS is defined as $Q_{\rm LSS}$. The results of the quantification of the large-scale environment for isolated, pair, and triplet galaxies are shown in Fig.~\ref{Fig:isolparam3}. 

For isolated pairs and isolated triplets, the tidal strength produced by galaxies of the same system, in comparison to the tidal strength exerted by galaxies in the LSS, leads to an evaluation of the effects of the isolated systems with respect to the large-scale environment. 
We therefore define $Q_{\rm{pair}}$ and $Q_{\rm{triplet}}$ for pairs and triplets, taking only the galaxies that belong to the same isolated system into account (the \emph{B} galaxy for pairs, and the \emph{B} and \emph{C} galaxies in the case of triplets). We also define $Q_{\rm{total}}$, which accounts for the total amount of tidal strength exerted by both the neighbour galaxies within the pair and triplet and by the galaxies from the large-scale environment. Therefore, the ratios $\frac{Q_{\rm{pair}}}{Q_{\rm{total}}}$ and $\frac{Q_{\rm{triplet}}}{Q_{\rm{total}}}$ take values from 0 to 1 if the galaxies of the isolated system amount to 0\% to 100\% of the total tidal strengths. These ratios for isolated pairs and isolated triplets are tabulated in Col. 8 in Tables~\ref{tab:tab2_2} and~\ref{tab:tab3_2}, respectively.

\begin{table}
  \caption{\label{tab:tab2_2}Isolated pairs: relation to their local and large-scale environments.} 
\centering
\begin{tabular}{cccccccc}
\hline \hline
(1) & (2) & (3) & (4) &  (5) & (6) & (7) & (8) \\
 SIP &   $k_{\rm{LSS}}$  &  $f_{\rm{LSS}}$  & $d_{\rm{NN}}$ &  $\eta_{k,\rm{LSS}}$  &  $Q_{\rm{LSS}}$  &  $Q_{\rm{pair}}$ &  $\frac{Q_{\rm{pair}}}{Q_{\rm{total}}}$ \\
  & &  & (Mpc) & & & &  \\
\hline
     1     &    11     &     1     &     1.57     &    -1.03     &    -5.34     &    -2.28     &         1.00    \\
     2     &    42     &     0     &     1.67     &    -1.20     &    -5.53     &    -3.50     &         0.99    \\
     3     &     9     &     1     &     1.32     &    -1.79     &    -4.80     &    -1.25     &         1.00    \\
     4     &    17     &     1     &     2.13     &    -1.39     &    -5.29     &     1.52     &         1.00    \\
     5     &    23     &     1     &     1.38     &    -1.34     &    -5.13     &    -2.08     &         1.00    \\
     6     &    43     &     1     &     1.55     &    -0.96     &    -3.92     &    -2.54     &         0.96    \\
     7     &    34     &     1     &     1.34     &    -0.63     &    -5.25     &    -3.66     &         0.97    \\
     8     &     4     &     1     &     1.71     &    -1.98     &    -5.40     &    -2.85     &         1.00    \\
     9     &    36     &     1     &     1.21     &    -1.47     &    -4.23     &    -1.61     &         1.00    \\
    10     &    29     &     1     &     1.15     &    -0.71     &    -4.37     &    -2.14     &         0.99    \\    
\ldots & \ldots & \ldots & \ldots & \ldots  & \ldots & \ldots & \ldots \\
\hline   
\end{tabular}
\tablefoot{The full table is available in electronic form at the CDS. 
The columns correspond to: 
(1) isolated pair identification; 
(2) $k_{\rm{LSS}}$, number of LSS associations; 
(3) $f_{\rm{LSS}}$, footprint flag: ``1'' if the \emph{A} galaxy has a 5\,Mpc field radius in the SDSS footprint, ``0'' if not; 
(4) $d_{\rm{NN}}$, distance from the \emph{A} galaxy to the first nearest neighbour of the LSS (from 1\,Mpc to 5\,Mpc);
(5) $\eta_{k,\rm{LSS}}$, projected density estimation of the LSS; 
(6) $Q_{\rm{LSS}}$, tidal strength estimation of the LSS; 
(7) $Q_{\rm{pair}}$, tidal strength estimation on the \emph{A} galaxy exerted by the \emph{B} galaxy; 
(8) $\frac{Q_{\rm{pair}}}{Q_{\rm{total}}}$, relation between $Q_{\rm{pair}}$ and the total tidal strengths on the \emph{A} galaxy $Q_{\rm{total}}$, from 0 to 1 if the isolated system amounts from 0\% to 100\% to the total tidal strength.}
\end{table}

\begin{table}
  \caption{\label{tab:tab3_2}Isolated triplets: relation with their local and large-scale environments.} 
\centering
\begin{tabular}{cccccccc}
\hline \hline
(1) & (2) & (3) & (4) &  (5) & (6) & (7) & (8)  \\
 SIT &  $k_{\rm{LSS}}$  & $f_{\rm{LSS}}$  &  $d_{\rm{NN}}$ &  $\eta_{k,\rm{LSS}}$  &  $Q_{\rm{LSS}}$  &  $Q_{\rm{triplet}}$ &  $\frac{Q_{\rm{triplet}}}{Q_{\rm{total}}}$ \\
    & & & (Mpc) & & & & \\
\hline
     1     &     9     &     1     &     2.53     &    -1.82     &    -5.87     &    -2.50     &         1.00    \\
     2     &     9     &     1     &     1.01     &    -1.43     &    -4.95     &    -1.49     &         1.00    \\
     3     &    10     &     1     &     1.50     &    -1.65     &    -5.68     &    -1.26     &         1.00    \\
     4     &     4     &     1     &     2.67     &    -2.21     &    -5.74     &    -2.34     &         1.00    \\
     5     &    13     &     1     &     1.95     &    -1.21     &    -5.55     &    -3.78     &         0.98    \\
     6     &    15     &     1     &     1.51     &    -0.93     &    -5.81     &    -0.92     &         1.00    \\
     7     &     9     &     1     &     1.44     &    -1.56     &    -5.12     &    -1.86     &         1.00    \\
     8     &    12     &     1     &     1.36     &    -0.96     &    -5.22     &    -2.91     &         1.00    \\
     9     &    36     &     0     &     1.10     &    -0.78     &    -5.13     &    -2.12     &         1.00    \\
    10     &    12     &     1     &     1.50     &    -1.14     &    -4.62     &    -0.27     &         1.00    \\    
    
\ldots & \ldots & \ldots & \ldots & \ldots  & \ldots & \ldots & \ldots \\
\hline   
\end{tabular}
\tablefoot{The full table is available in electronic form at the CDS. 
The columns correspond to: 
(1) isolated triplet identification; 
(2) $k_{\rm{LSS}}$, number of LSS associations; 
(3) $f_{\rm{LSS}}$, footprint flag: ``1'' if the \emph{A} galaxy has a 5\,Mpc field radius in the SDSS footprint, ``0'' if not; 
(4) $d_{\rm{NN}}$, distance from the \emph{A} galaxy to the first nearest neighbour of the LSS (from 1\,Mpc to 5\,Mpc); 
(5) $\eta_{k,\rm{LSS}}$, projected density estimation of the LSS; 
(6) $Q_{\rm{LSS}}$, tidal strength estimation of the LSS;  
(7) $Q_{\rm{triplet}}$, tidal strength estimation on the A exerted by the \emph{B} and \emph{C} galaxies; 
(8) $\frac{Q_{\rm{triplet}}}{Q_{\rm{total}}}$, relation between $Q_{\rm{triplet}}$ and the total tidal strengths on the \emph{A} galaxy $Q_{\rm{total}}$, from 0 to 1 if the isolated system amounts from 0\% to 100\% to the total tidal strength.}
\end{table}


\section{Discussion} \label{Sec:discussion3}

\subsection{The catalogues} \label{Sec:dis_cat3}

To study the intrinsic processes that affect galaxy evolution, it is crucial to have a reference sample of galaxies where the effects of their local and large-scale environments are minimised and quantified.
As shown in Sect.~\ref{Sec:res_cat3}, the catalogues of isolated galaxies, isolated pairs, and isolated triplets have been homogeneously selected using an automated method. The SIG catalogue is composed of 3~702 galaxies with no neighbours within $\Delta\,\varv\,\leq\,500$\,km\,s$^{-1}$ in a field radius of 1\,Mpc. In the case of the SIP and SIT catalogues, we identify 1~240 isolated pairs and 315 isolated triplets, with no neighbours within $\Delta\,\varv\,\leq\,500$\,km\,s$^{-1}$ in a field radius of 1\,Mpc. These uniformly selected catalogues offer the opportunity to test model predictions for galaxy evolution. 

The velocity difference of pairs and triplets shown in Fig.~\ref{Fig:physicals3} follows a Gaussian distribution, meaning that most of the pairs and triplets are relaxed systems. Indeed, if the pairs and triplets were mostly fly-by encounters, there would be no preferred values for the velocity differences between the central galaxies and their companions. A flat distribution would be observed, as is the case for $\Delta\,\varv\,\gtrsim\,300$\,km\,s$^{-1}$. As a result, the Gaussian distribution shows that most of the pair and triplet systems are already virialised and suggests that they did not form by chance unless the captures occurred at a primordial epoch. These pair and triplet samples should then shed light on the effects of secular evolution of galaxies in interaction with respect to isolated galaxies or galaxies just temporarily affected by some fly-by companions. It is left for simulations to show if the most isolated systems today are likely to have spent all or most of their lives in isolation. For each galaxy in the catalogues, we provide its position, redshift, and quantification of the environment (see Table~\ref{tab:tab1} for isolated galaxies, Tables~\ref{tab:tab2} and \ref{tab:tab2_2} for isolated pairs, and Tables~\ref{tab:tab3} and \ref{tab:tab3_2} for isolated triplets). 

\begin{figure}
\centering
\includegraphics[width=\columnwidth]{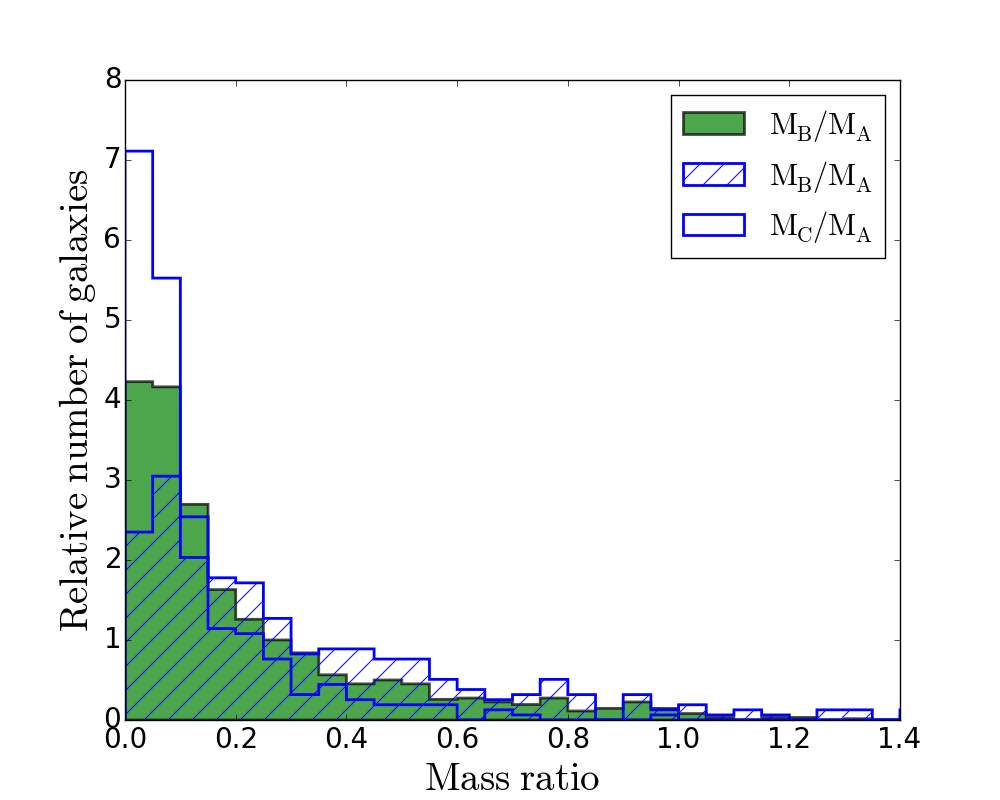} \\
\includegraphics[width=\columnwidth]{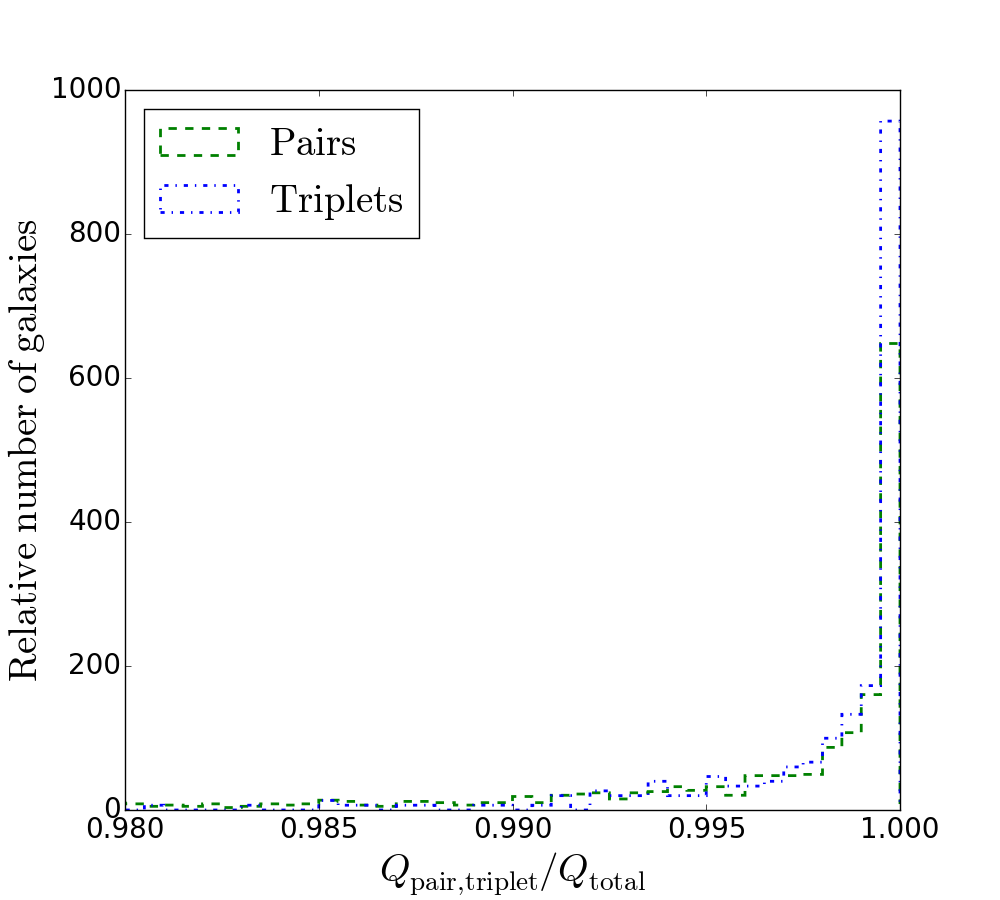}
\caption{{\it (Upper panel):} Distribution of the mass ratio between the \emph{A} galaxy and the \emph{B} galaxy in isolated pairs (green filled histogram), between the \emph{A}  and \emph{B} galaxies for isolated triplets (blue unfilled histogram), and between the \emph{A} galaxy and the \emph{C} galaxy for isolated triplets (blue hatched histogram). {\it (Lower panel):} Distribution of the ratios $\frac{Q_{\rm{pair}}}{Q_{\rm{total}}}$ (green dashed histogram) and $\frac{Q_{\rm{triplet}}}{Q_{\rm{total}}}$ (blue dash-dotted histogram) for isolated pairs and isolated triplets, respectively.}
\label{Fig:properties3}
\end{figure}

In the upper panel in Fig.~\ref{Fig:properties3}, we show the mass ratio for galaxies in the catalogues of isolated pairs and isolated triplets. In isolated pairs, the mass of the \emph{B} galaxy is typically one-tenth of the mass of the \emph{A} galaxy. This means that isolated pairs usually show a hierarchical structure, therefore it is less probable to find a pair with two similar mass galaxies. For the triplets, the mass ratio of the \emph{A} and \emph{C} galaxies is much smaller ($\sim$\,1/100) than the \emph{A} and \emph{B} galaxies. This result is also visible in some of the examples in Fig.~\ref{Fig:charts10triplets}, the \emph{C} galaxy is usually very much smaller than the \emph{A} and \emph{B} galaxies, and it also presents a hierarchical structure. For isolated pairs and isolated triplets, we consider the  central (\emph{A}) galaxy to be the brightest, which is not directly translated into its being the most massive galaxy, so we find a small fraction of mass ratios higher than one in the upper panel of Fig.~\ref{Fig:properties3}.

Our results are compatible with \citet{2012MNRAS.422.2187M,2013MNRAS.429..792M} and, more recently, with \citet{2014MNRAS.442..347R}, who characterised the population of satellites around massive ellipticals ($M_\star~\gtrsim~2~\times~10^{11}\,M_\odot$). They find that on average, there is  about one satellite around each central galaxy, down to a mass ratio 1/100. In particular, only about 23\% of the satellites have mass ratios around 1/10, versus $\sim$\,50\% for mass ratios 1/100, and even $>$\,60-70\% down to 1/400. Even if we are restricted to lower mass central galaxies (8~$\lesssim$~log($M_\star$)~$\lesssim$~11.5), we find that the number of physically bound pairs and triplets increases from a mass ratio 1/10 to mass ratio down to 1/100.

\subsection{Comparison with previous existing catalogues} \label{Sec:dis_others}

As described in Sect.~\ref{Sec:res_others}, we selected the CIG and the UNAM-KIAS to compare with the SIG catalogue, the KPG and the sample of close pairs compiled by \citet{2008ApJ...685..235P} to compare with the SIP catalogue, and the KTG and the catalogue of triplets compiled by \citet{2012MNRAS.421.1897O} to compare with the SIT catalogue. 

The results of the comparison with the CIG, KPG, and KTG are shown in the first three rows of Table~\ref{tab:comparison}. The small overlap between the initial number of galaxies in the comparison samples and the number of galaxies in the primary sample is mainly due to the different sky coverage of the spectroscopic SDSS and the POSS (Palomar Observatory Sky Survey). As shown in Table~\ref{tab:comparison}, about 29\% of the CIG galaxies are in common with our catalogue of isolated galaxies, and only about 2\% of the KPG galaxies and 5\% of KTG triplets are found in our catalogues of isolated pairs and isolated triplets, respectively. These results agree with \citet{2013A&A...560A...9A}, who find that about 25\% of the CIG galaxies (105 of the 411 CIG galaxies considered in their study) have no known neighbours with $|\Delta\,\varv|~\leq~500$\,km\,s$^{-1}$ within 1\,Mpc in the SDSS-DR9. The visual isolation criterion based on apparent magnitude differences (due to the small fraction of galaxies with known redshifts at this time) to compile the CIG (and similarly to compile the KPG and KTG) misses close, small neighbours with similar redshift. Moreover, it introduces false pairs and triplets with a very high velocity difference between their members due to projection effects.

\begin{figure}
\centering
\includegraphics[width=\columnwidth]{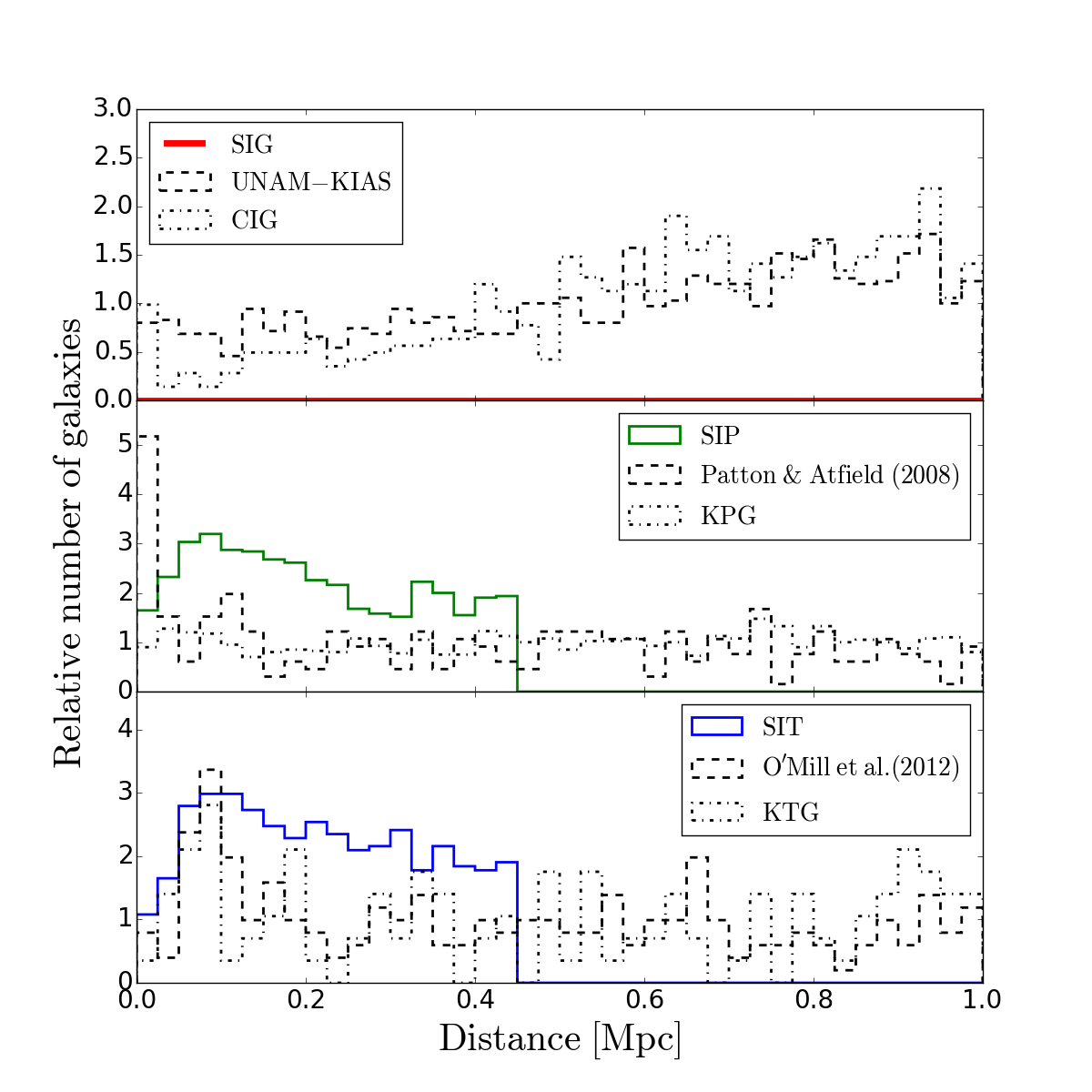}
\caption[Comparison samples]{Normalised distribution of the distance (in Mpc) for neighbours with $\Delta\,\varv\,\leq\,500$\,km\,s$^{-1}$ within 1\,Mpc around galaxies in the comparison samples. 
{\it (Upper panel):} The red solid histogram represents neighbours around our isolated galaxies is zero by definition. The black dashed and black dash-dotted histograms correspond to the distributions for neighbours around galaxies in the UNAM-KIAS and the CIG, respectively. 
{\it (Middle panel):} Distributions for neighbours around galaxies in the KPG (black dash-dotted histogram) and the catalogue of \citet{2008ApJ...685..235P} (black dashed histogram). The green solid histogram corresponds to the distribution of the distance for the \emph{B} galaxies in our isolated pairs. 
{\it (Lower panel):} Distributions for neighbours around galaxies in the KTG (black dash-dotted histogram) and the catalogue of \citet{2012MNRAS.421.1897O} (black dashed histogram). The blue solid histogram corresponds to the distribution of the distance for the \emph{B} and \emph{C} galaxies in our isolated triplets.}
\label{Fig:comparison}
\end{figure}

The three last rows of Table~\ref{tab:comparison} show the result of the comparison with the catalogues based on SDSS spectroscopy. To help the interpretation of the results, we look for neighbours with $\Delta\,\varv\,\leq\,500$\,km\,s$^{-1}$ within 1\,Mpc around each galaxy in the comparison samples. Figure~\ref{Fig:comparison} shows the normalised distribution of distance for the similar-redshift neighbours with respect to the central galaxy. 

We find 43.5\% of the UNAM-KIAS galaxies in common with our isolated galaxies. This average overlap is mainly due to the different isolation criteria. The isolation criterion followed by \citet{2010AJ....139.2525H} is a mixed criterion between the CIG and a pure 3D definition. The distribution of the distance for neighbours with $\Delta\,\varv\,\leq\,500$\,km\,s$^{-1}$ within 1\,Mpc around galaxies in the comparison samples is shown in Fig.~\ref{Fig:comparison}. As shown in the upper panel of the figure, there are close neighbours with similar redshift missed by the UNAM-KIAS isolation criterion (similarly for CIG galaxies) and, by definition, zero neighbours for our isolated galaxies. In this sense, the SIG represents the first public catalogue of isolated galaxies based on a pure 3D isolation definition.

Regarding pairs and triplets, there is a very small overlap between the number of galaxies in our primary samples and the initial number of galaxies in the samples compiled by \citet{2008ApJ...685..235P}, because 90\% of their central galaxies have $m_{r} > 15.7$, and by \citet{2012MNRAS.421.1897O} since 97\% of their central galaxies have $z > 0.08$. The middle and lower panels of Fig.~\ref{Fig:comparison} show that the close pairs and triplets found in the comparison samples are less isolated than our isolated pairs and isolated triplets. (The systems have neighbour galaxies, at least up to 1\,Mpc, that may influence their evolution.) In fact only 7\% of the isolated pairs and 5\% of the isolated triplets are in common. An isolation criteria is not used in these samples, so our isolated pairs and isolated triplets are not as close as theirs but are very well-isolated systems.

To date, there are no public catalogues that allow a direct separation between the effects of one-on-one interactions and the effects of the large-scale environment when studying galaxy properties. Galaxies in our catalogues have been homogeneously selected under the same isolation criterion, so that the SIG catalogue is complemented by the SIP and SIT catalogues for testing galaxy evolution and secular processes. Our catalogues of isolated galaxies, isolated pairs, and isolated triplets represent suitable samples of isolated systems for studying the link of galaxy properties with their environment for bright galaxies in the local Universe. 

\subsection{Relation with the large-scale structure} \label{Sec:dis_LSS}

\begin{figure}
\begin{center}
\includegraphics[width=.5\textwidth]{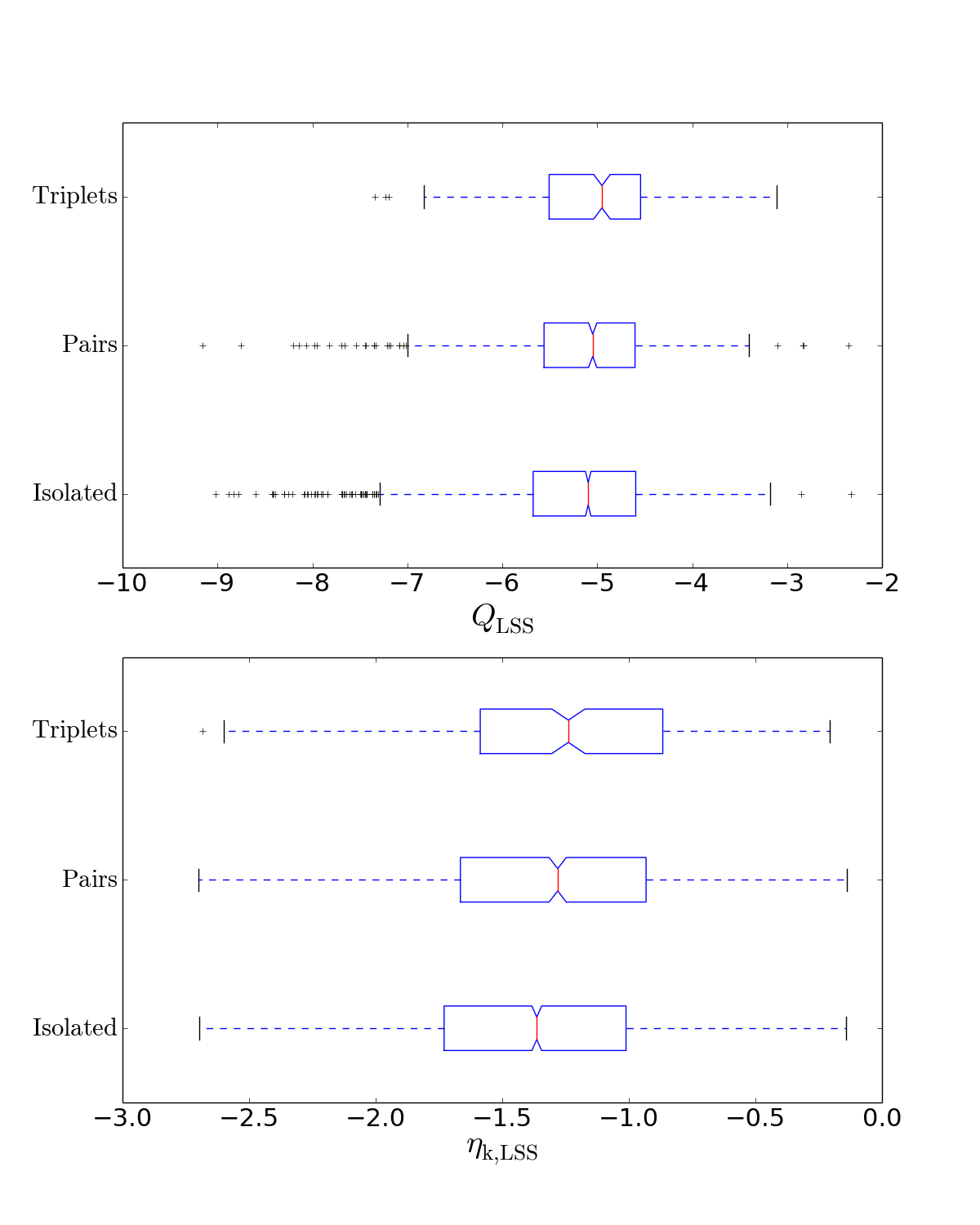}  
\end{center}
\caption{Comparison of LSS isolation parameters for isolated galaxies, isolated pairs, and isolated triplets. The main body of the boxplot shows the interquartile range (IQR) of the median (represented by the red vertical line) and its 95\% confidence intervals; meanwhile, dashed lines correspond to 1.5$\times$IQR and the outliers are represented by black pluses. {\it (Upper panel):} Boxplot representation of the distribution of the tidal strength $Q_{\rm LSS}$ for isolated galaxies (lower boxplot), isolated pairs (middle boxplot), and isolated triplets (upper boxplot). {\it (Lower panel):} Boxplot representation of the distribution of the projected density $\eta_{k,\rm LSS}$ for isolated galaxies, isolated pairs, and isolated triplets following the same scheme.} \label{Fig:boxplot}
\end{figure}

As introduced in Sect.~\ref{Sec:quantification3}, we used the projected density $\eta_{k,\rm{LSS}}$ and tidal strength $Q_{\rm{LSS}}$ to quantify the large scale environment for isolated galaxies, isolated pairs, and isolated triplets. The upper panels in Fig.~\ref{Fig:isolparam3} show that there is no difference in their relation with the LSS between the three samples. The mean values of the tidal strength for isolated galaxies, isolated pairs, and isolated triplets are $-5.09\pm 0.03$, $-5.05\pm 0.04$, and $-4.95\pm 0.09$, respectively. (Uncertainties are given by the 95\% confidence interval of the median.) Even if the mean value of the projected density for isolated galaxies ($-1.38\pm 0.02$) is slightly lower than for isolated pairs and isolated triplets ($-1.28\pm 0.03$ and $-1.24\pm 0.07$, respectively), the boxplots of Fig.~\ref{Fig:boxplot} show that there is no significant difference between the three populations. This means that the isolated galaxies are as isolated, with respect to the large scale environment, as the isolated pairs and triplets. This suggests that these systems have a common origin in their formation and evolution.

When taking only the physical associations into account of the galaxies in pairs (B) and triplets (B, C), which define the tidal strengths $Q_{\rm{pair}}$ and $Q_{\rm{triplet}}$, respectively, we see that their effects largely dominate (about 99.9\%) the tidal strengths generated by the LSS (see also lower panel in Fig.~\ref{Fig:properties3}). The tidal strengths $Q_{\rm{pair}}$ and $Q_{\rm{triplet}}$ are two orders of magnitude higher than the $Q_{\rm{LSS}}$ (see the lower right panel in Fig.~\ref{Fig:isolparam3}). Therefore, the effects of the LSS are negligible in contrast to the effect of one-on-one interactions.

\begin{figure*}  
\begin{center}
\includegraphics[width=\textwidth]{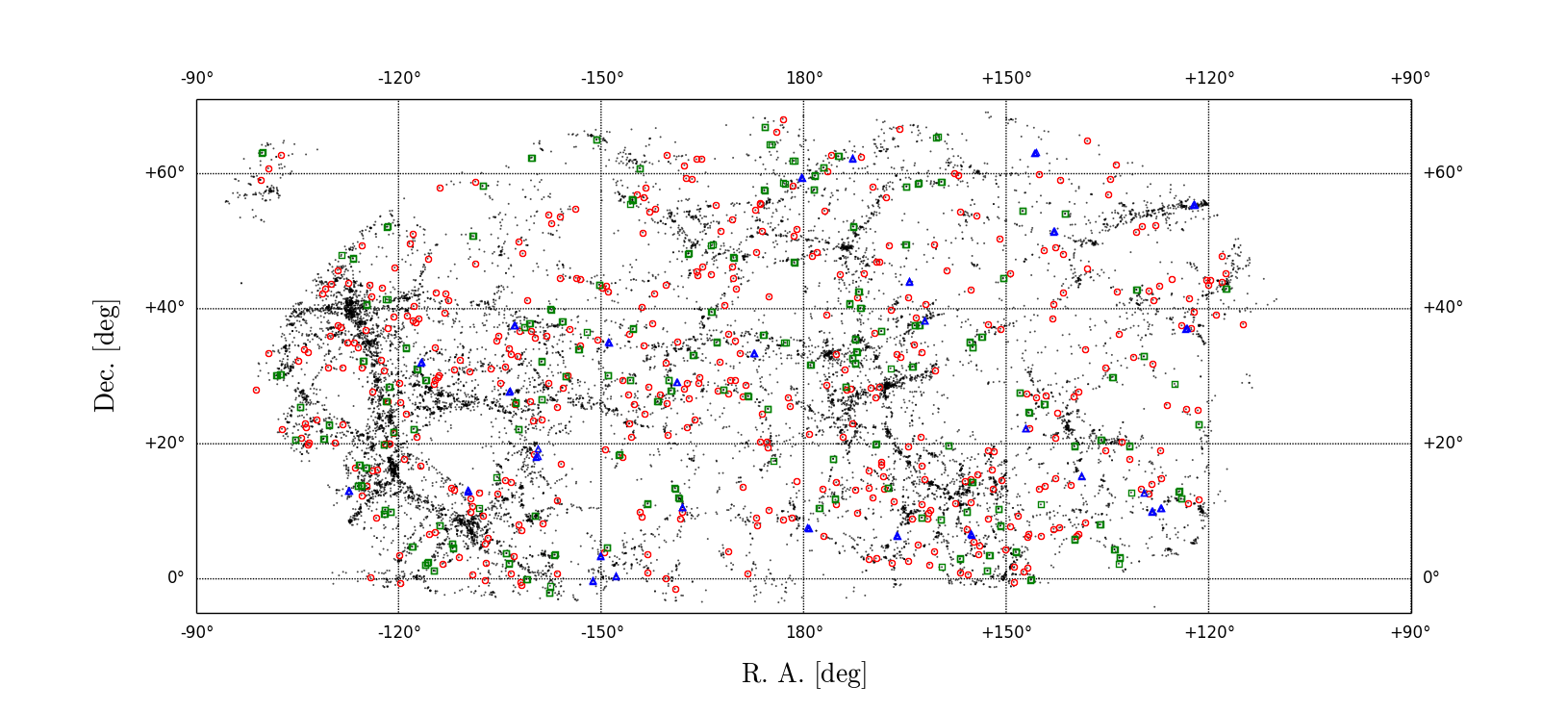} \\
\includegraphics[width=.65\textwidth]{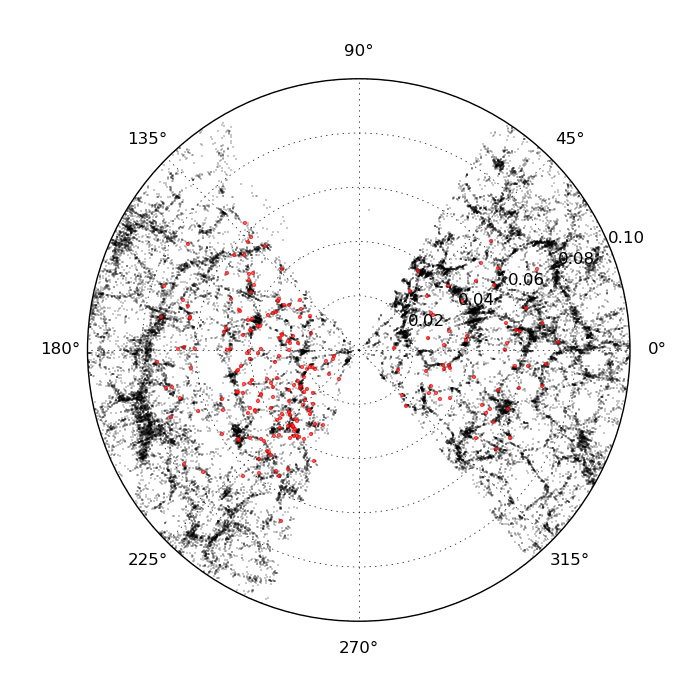}
\end{center}
\caption[Distribution of isolated galaxies with respect to the LSS]{{\it (Upper panel):} Equidistant cylindrical projection of the LSS for galaxies in the Northern Galactic Cap in the redshift range $0.030 < z < 0.035$ (North is up, east to the left). Red circles, green squares, and blue triangles represent isolated galaxies, isolated pairs, and isolated triplets, respectively. {\it (Lower panel):} Pie chart projection of the LSS for the same galaxies within $-2$ and $2$ degrees in declination. The over-plotted red disks represent isolated galaxies within the same declination range.} \label{Fig:alumnosLSS1}
\end{figure*}

\begin{figure*}  
\begin{center}
\includegraphics[width=.65\textwidth]{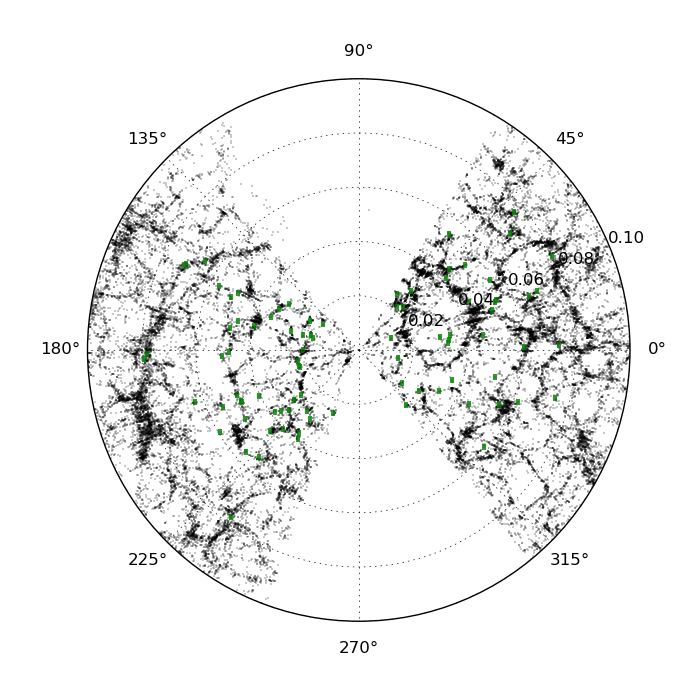} \\
\includegraphics[width=.65\textwidth]{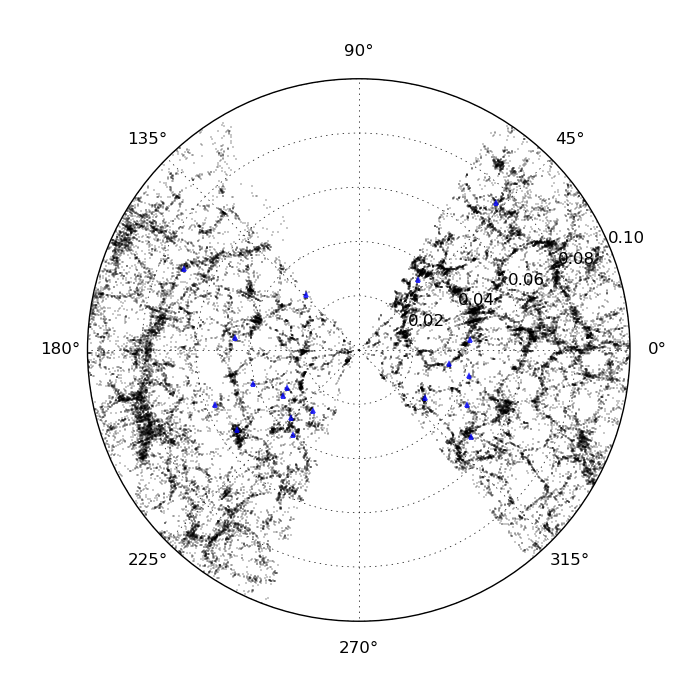} 
\end{center}
\caption{Distribution of isolated pairs and isolated triplets in the LSS. Same as the lower panel in Fig.~\ref{Fig:alumnosLSS1} but for isolated pairs (green squares in the upper panel) and isolated triplets (blue triangles in the lower panel).} \label{Fig:alumnosLSS2}
\end{figure*}

We find very few galaxies with no neighbours within $\Delta\,\varv\,\leq\,500$\,km\,s$^{-1}$ up to 5\,Mpc. One can expect that these galaxies would be located in voids. The smallest identifiable voids in the local Universe have radii $\sim 7$\,h$^{-1}$\,Mpc \citep{2006ApJ...653..969T}, with voids usually having characteristic radii of 10-40\,h$^{-1}$\,Mpc \citep{2012ApJ...761...44S}.
Nevertheless, Figs.~\ref{Fig:alumnosLSS1} and~\ref{Fig:alumnosLSS2} (see also Appendix~\ref{Sec:App1} for details) show that in general, most of the isolated galaxies, isolated pairs, and isolated triplets belong to the outer parts of filaments, walls, and clusters, and that they generally differ from the void population of galaxies.

To further explore this difference with void galaxies, we compared our catalogue of isolated galaxies with the public catalogue of voids compiled by \citet{2012MNRAS.421..926P}, based on SDSS spectroscopy. We find that 33\% of our isolated galaxies are located in voids. This result is expected since the relation of the radius where isolation is defined (1\,Mpc) with respect to the typical size of regular clusters (from 1 to 10\,Mpc) is much smaller than the relation with the typical radius of voids in the Universe (24\,Mpc). Additionally, we found that only 14\% of void galaxies, within the same limited volume as our primary sample are in isolation according to our criteria. This result reinforces the argument of the different natures of void and isolated galaxies.

\section{Summary and conclusions} \label{Sec:con3}

We present the SIG, SIP, and SIT catalogues of isolated galaxies, isolated pairs, and isolated triplets in the local Universe ($z\leq0.08$ and $m_r\leq15.7$\,mag). They were automatically and uniformly selected using the SDSS-DR10. These systems show no known neighbours within $\Delta\,\varv\,\leq\,500$\,km\,s$^{-1}$ and radius $R=1$\,Mpc. All the systems in the three final catalogues were visually inspected. The catalogues, with galaxy positions, redshifts, and degrees of relation with their physical and LSS environments, are publicly available to the scientific community. In addition, we provide interactive tools for visualising the relation of the galaxies in the samples with the LSS or any other catalogue.

The relations of the isolated galaxies, isolated pairs, and isolated triplets with the LSS of the local Universe are characterised using the projected density $\eta_{k,\rm{LSS}}$ and the tidal strength $Q_{\rm{LSS}}$ produced by the neighbours within $\Delta\,\varv\,\leq\,500$\,km\,s$^{-1}$ up to 5\,Mpc.

Our main conclusions are the following:

\begin{enumerate}
\item The SIG catalogue of isolated galaxies is composed of 3~702 galaxies with no neighbours within $\Delta\,\varv~\leq~500$\,km\,s$^{-1}$ in a field radius of 1\,Mpc. The sample represents about 11\% of the total number of galaxies in the local Universe considered in the study.

\item When analysing the distribution in velocity difference $\Delta\,\varv$ and projected distance $d$ space for isolated pairs and isolated triplet candidates, we find evidence of physically bound galaxies at distances $d~\leq~450$\,kpc within $\Delta\,\varv~\leq~160$\,km\,s$^{-1}$.

\item The SIP catalogue of isolated pairs contains 1~240 pairs, which were selected using the requirement that the faintest galaxy should be within $\Delta\,\varv~\leq~160$\,km\,s$^{-1}$ and lie at a distance $d~\leq~450$\,kpc with respect to the brightest one. The sample of isolated pairs represents about 7\% of the total number of galaxies in the local Universe considered in the study.

\item Similarly, the SIT catalogue of isolated triplets consists of 315 triplets, selected under the condition that the two faintest galaxies satisfy $\Delta\,\varv~\leq~160$\,km\,s$^{-1}$ and lie within $d~\leq~450$\,kpc of the brightest galaxy. The sample of isolated triplets represents about 3\% of the total number of galaxies in the local Universe considered in the study.

\item To reach a more homogeneous characterisation of the environment around the isolated samples, we analysed their large-scale environment up to 5\,Mpc. We found 31 very isolated galaxies, 9 isolated pairs, and no isolated triplet without any neighbour within $\Delta\,\varv\,\leq\,500$\,km\,s$^{-1}$ and up to 5\,Mpc. For the remaining primary galaxies in the three samples, we defined the projected density $\eta_{k,\rm{LSS}}$ and the tidal strength $Q_{\rm LSS}$ to quantify their isolation degree with respect to their large-scale environment.

\item The physical companions in pairs and triplets are the dominant factor in the tidal strengths $Q_{\rm pair}$ and $Q_{\rm triplet}$ exerted on the primary galaxies despite their hierarchical structure. This local tidal strength due to the physical companions is two orders of magnitude higher than the tidal strength due to the LSS.

\item There is no difference in the degree of interaction with the LSS for isolated singles, pairs, and triplets, which may suggest that they have a common origin (i.e., formation and evolution). We find that most belong to the outer parts of filaments, walls, and clusters, and generally differ from the void population of galaxies.
\end{enumerate}

In comparison with the existing literature, the SIG, SIP, and SIT catalogues of isolated galaxies, isolated pairs, and isolated triplets represent particularly useful samples for studies of both one-on-one and the large-scale effects on galaxy properties.

\begin{acknowledgements}

This work was partially supported by MICINN of Spain via grants AYA2011-24728, and from the ``Junta de Andalucía'' local government through the FQM-108 project.

This work was partially supported by Grant AYA2011-30491-C02-01, co-financed by MICINN and FEDER funds, and the Junta de Andalucia (Spain) grants P08-FQM-4205 and TIC-114, as well as under the EU 7th Framework Programme in the area of Digital Libraries and Digital Preservation (ICT-2009.4.1) Project reference: 270192.

MAF is also grateful for financial support from PIFI (funded by Chinese Academy of Sciences President's International Fellowship Initiative) Grant No. 2015PM056. This work was partly supported by the Strategic Priority Research Programme ``The Emergence of Cosmological Structures'' of the Chinese Academy of Sciences (CAS; grant XDB09030200), the National Natural Science Foundation of China (NSFC) with the Project Number of 11433003 and the ``973 Programme'' 2014 CB845705.

Funding for SDSS-III was provided by the Alfred P. Sloan Foundation, the Participating Institutions, the National Science Foundation, and the U.S. Department of Energy Office of Science. The SDSS-III web site is http://www.sdss3.org/. SDSS-III is managed by the Astrophysical Research Consortium for the Participating Institutions of the SDSS-III Collaboration including the University of Arizona, the Brazilian Participation Group, Brookhaven National Laboratory, University of Cambridge, University of Florida, the French Participation Group, the German Participation Group, the Instituto de Astrofisica de Canarias, the Michigan State/Notre Dame/JINA Participation Group, Johns Hopkins University, Lawrence Berkeley National Laboratory, Max Planck Institute for Astrophysics, New Mexico State University, New York University, Ohio State University, Pennsylvania State University, University of Portsmouth, Princeton University, the Spanish Participation Group, University of Tokyo, University of Utah, Vanderbilt University, University of Virginia, University of Washington, and Yale University.

This research made use of python ({\tt http://www.python.org}), of Matplotlib \citep{Hunter:2007}, a suite of open-source python modules that provides a framework for creating scientific plots, and Astropy, a community-developed core Python package for Astronomy \citep{2013A&A...558A..33A}. This research made use of the NASA/IPAC Extragalactic Database (NED), which is operated by the Jet Propulsion Laboratory, California Institute of Technology, under contract with the National Aeronautics and Space Administration. This research made use of NASA's Astrophysics Data System Bibliographic Services. We also acknowledge the use of STILTS and TOPCAT tools \citep{2005ASPC..347...29T}.

\end{acknowledgements}

\bibliography{astroph}

\begin{appendix}

\section{Data visualisation tools} \label{Sec:App1}

We observe that most of the isolated galaxies, isolated pairs, and isolated triplets belong to the outer parts of filaments, walls, and clusters, and generally differ from the void population of galaxies. To visualise this, we have developed interactive visualisation tools based on open source code. The visualisation tools allow one to compare the 3D positions of a sample (or samples) of isolated systems with respect to the locations of the LSS galaxies in their local and/or large scale environments. The code is available at \texttt{https://github.com/margudo/LSSGALPY}.

The basic functionality of these interactive tools is the use of different projections in the 3D space (right ascension, declination, and redshift) to study the relation of the galaxies with the LSS. In particular, we use a Mollweide projection (the code also works with tens of other types of projections) in combination with a wedge diagram, and viceversa. Therefore, we can visualise the locations of the galaxies in our study for different values of redshifts and redshift ranges (by moving the blue bars displayed under the sky map in Fig.~\ref{Fig:tool1}). Similarly, for different values of the declinations and declination ranges (blue bars below wedge diagram in Fig.~\ref{Fig:tool2}), one can visualise how the isolated galaxies, isolated pairs, and isolated triplets are related to the galaxies in the LSS. 

These tools were tested using up to 30 million objects, and they still work perfectly and smoothly on any standard laptop. The whole code is 100\% based on free software (MIT License), making extensive use of the Python language. We have developed the code under a Linux platform, but it may be straightforwardly ported to any other operating system (Windows, Mac, FreeBSD, etc.).

\begin{landscape}
\begin{center}
\begin{figure}
\includegraphics[width=1.35\textwidth]{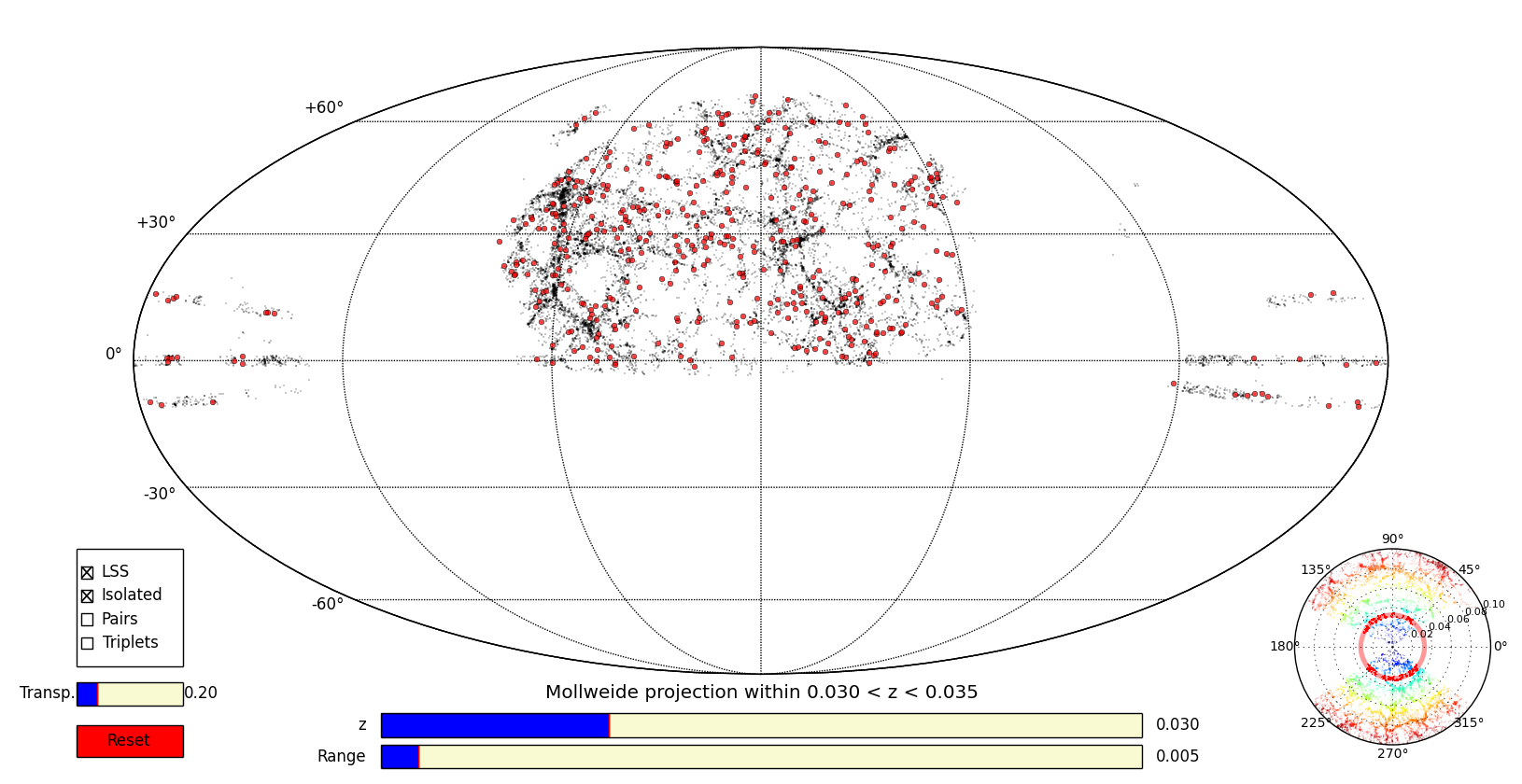}
\caption{Interactive 3D visualisation software: Mollweide projection. Mollweide projection of the LSS for galaxies (black points) in the redshift range $0.030 < z < 0.035$ as shown in the blue bars in the lower part of the figure (default values). Red disks represent isolated galaxies within the same redshift range. Isolated pairs and isolated triplets can be shown by selecting the samples in the sample box (panel in the left part of the figure). The transparency of the represented samples can be modified by sweeping the blue bar under the samples box. To guide the eye, a wedge diagram, for LSS galaxies within -2 and 2 degrees in declination, is shown in the right lower part of the figure. Colour code according to the redshift from $z=0$ (blue) to $z=0.10$ (red). The red ring in the polar representation corresponds to the selected redshift range in the central Mollweide projection. One can return to the default values with the reset button.}
\label{Fig:tool1}
\end{figure}
\end{center}
\end{landscape}

\begin{landscape}
\begin{center}
\begin{figure}
\includegraphics[width=1.35\textwidth]{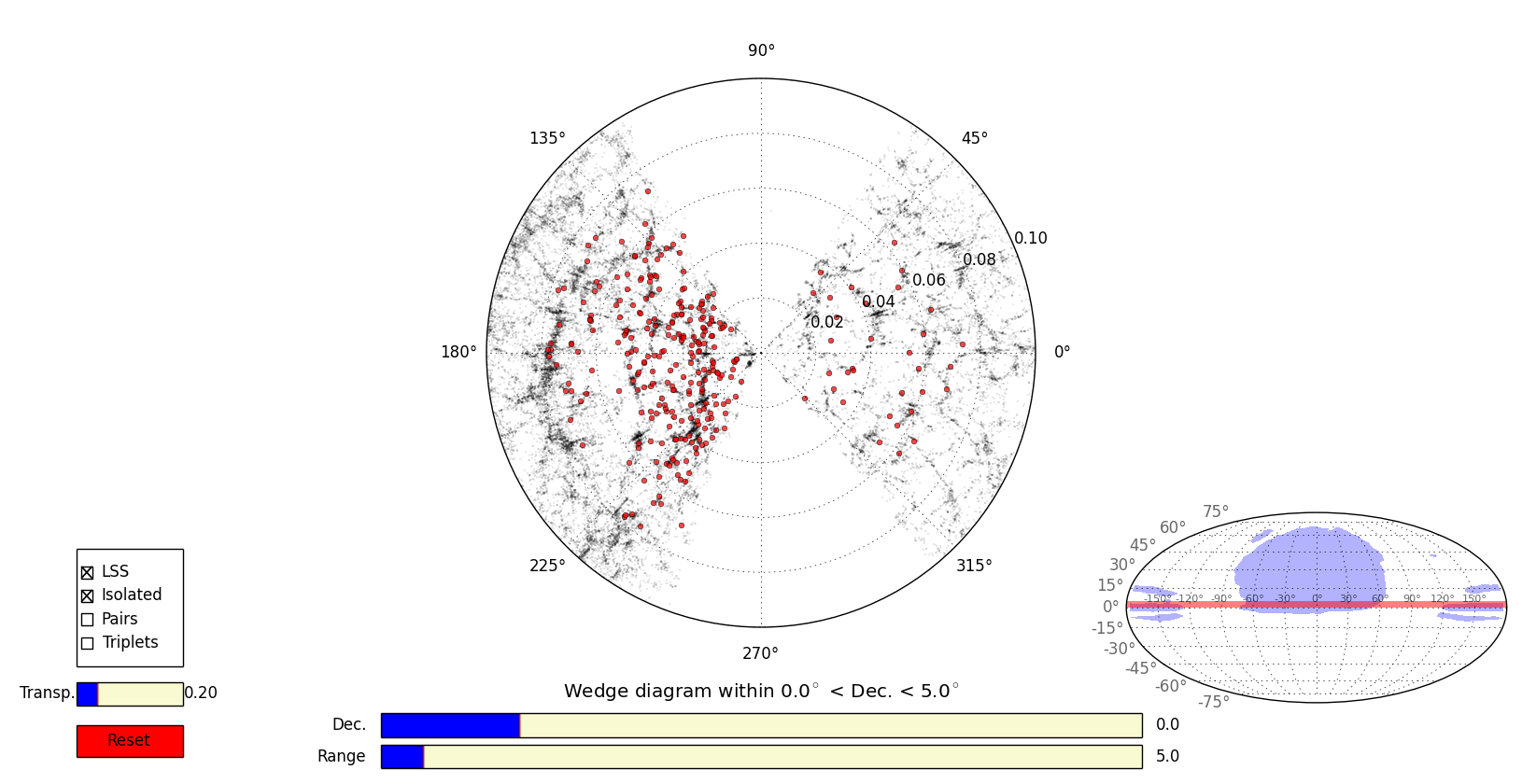}
\caption{Interactive 3D visualisation software: wedge diagram. Wedge diagram for LSS galaxies (black points) within 0 and 5 degrees in declination as shown in the blue bars in the lower part of the figure (default values). As in Fig.~\ref{Fig:tool1} red circles represent isolated galaxies within the same declination range. Isolated pairs and isolated triplets can be shown by selecting them in the samples box. To guide the eye, a Mollweide projection, for LSS galaxies in the redshift range $0.030 < z < 0.035$, is shown in the right lower part of the figure. The red stripe in the Mollweide projection corresponds to the selected declination range in the central wedge diagram. One can return to the default values with the reset button.}
\label{Fig:tool2}
\end{figure}
\end{center}
\end{landscape}

\section{Images}

We show some images for galaxies in the SIG, SIP, and SIT catalogues.

\begin{figure*}
\begin{center}
\includegraphics[width=0.27\textwidth]{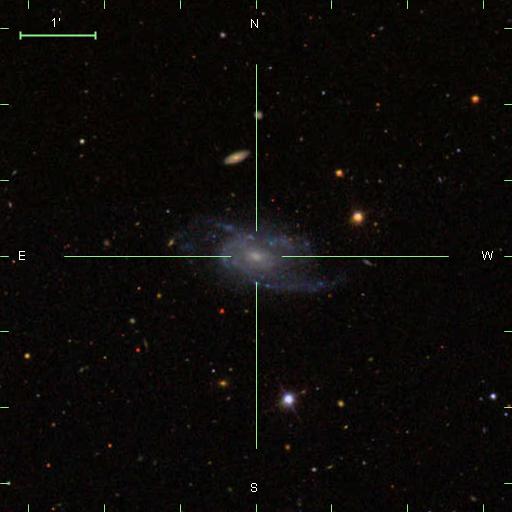}
\includegraphics[width=0.27\textwidth]{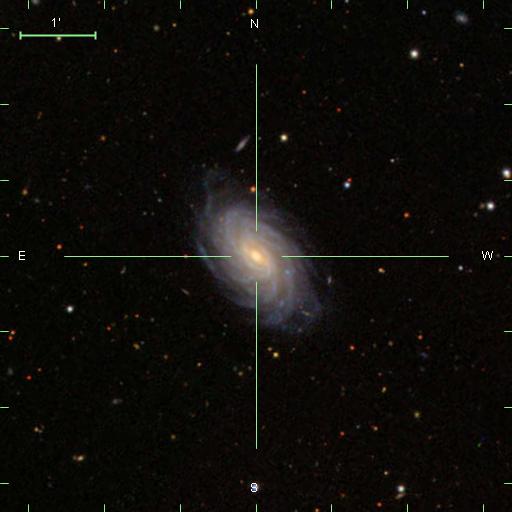}
\includegraphics[width=0.27\textwidth]{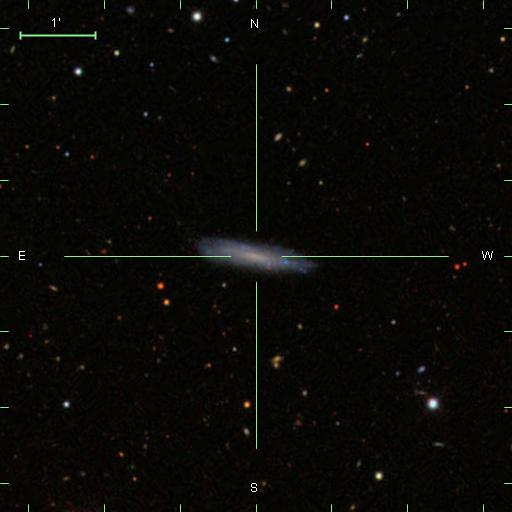} \\
\includegraphics[width=0.27\textwidth]{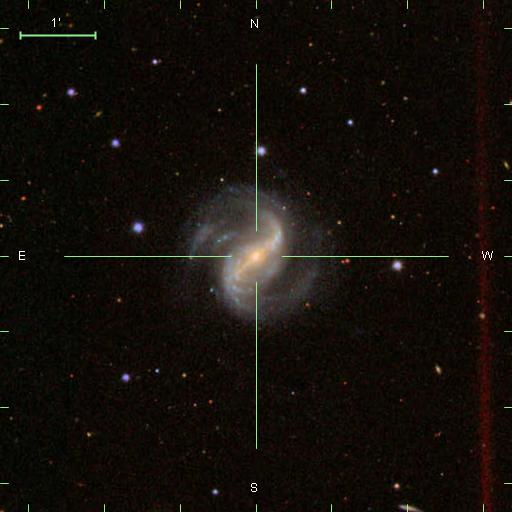}
\includegraphics[width=0.27\textwidth]{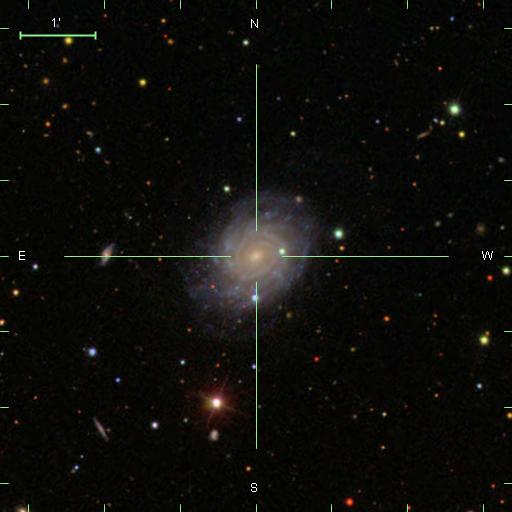}
\includegraphics[width=0.27\textwidth]{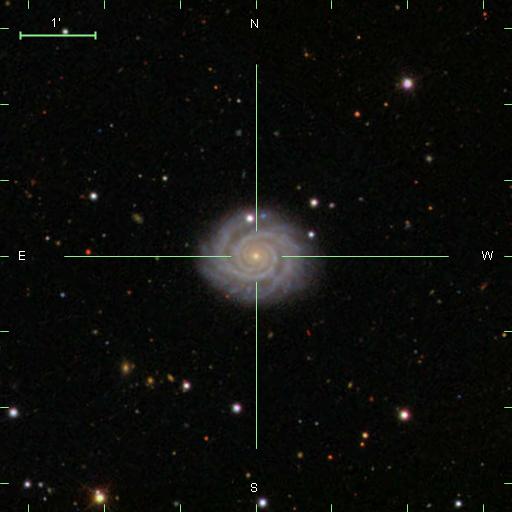} \\
\includegraphics[width=0.27\textwidth]{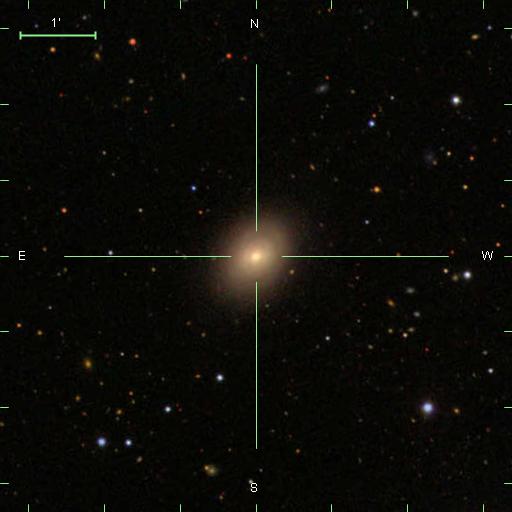}
\includegraphics[width=0.27\textwidth]{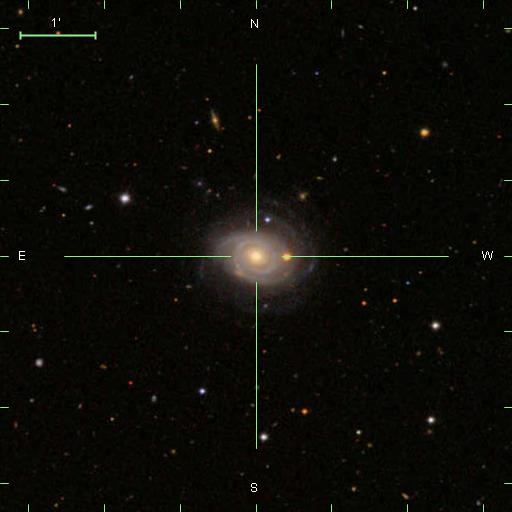}
\includegraphics[width=0.27\textwidth]{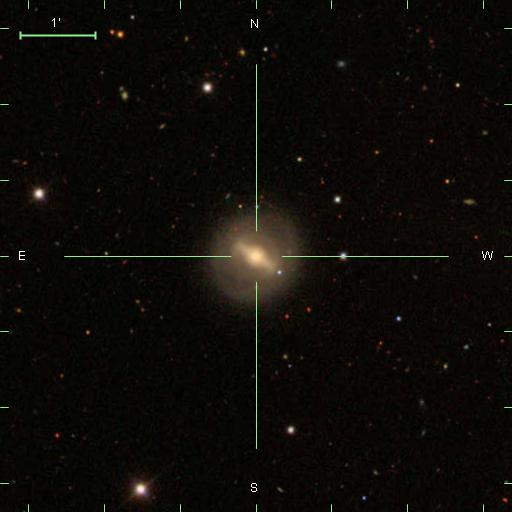}
\end{center}
\caption[Example images for isolated galaxies]{SDSS three-colour images (field of view of 6\farcm7, North is up, east is to the left) of nine isolated galaxies in the SDSS-DR10 footprint. {\it From upper left to lower right:} SIG 2053, 1947, 950, 1144, 2719, 931, 2459, 3061, and 3289.}\label{Fig:charts10isol}
\end{figure*}

\begin{figure*}
\begin{center}
\includegraphics[width=0.27\textwidth]{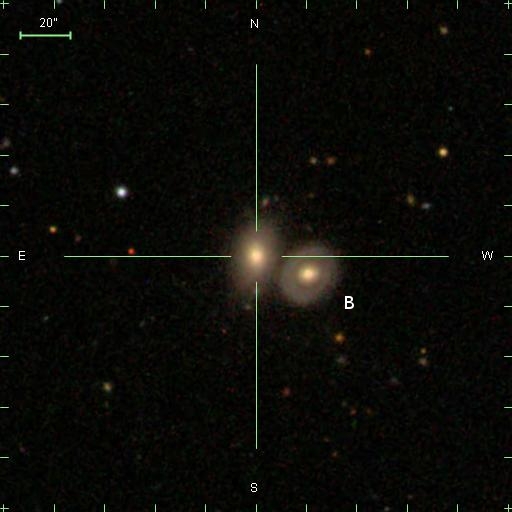}
\includegraphics[width=0.27\textwidth]{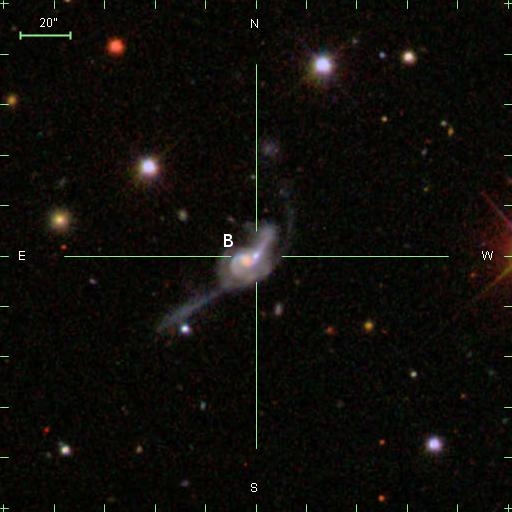}
\includegraphics[width=0.27\textwidth]{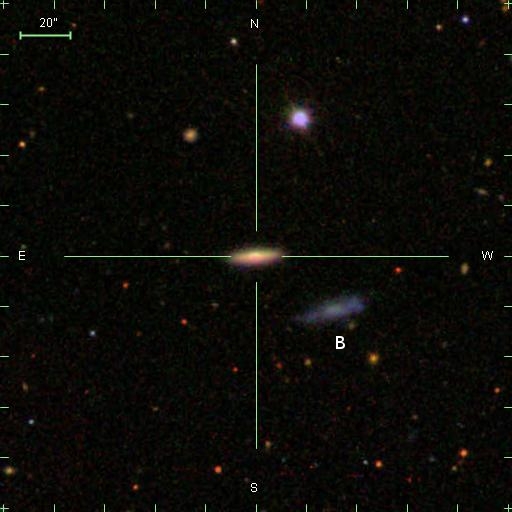} \\
\includegraphics[width=0.27\textwidth]{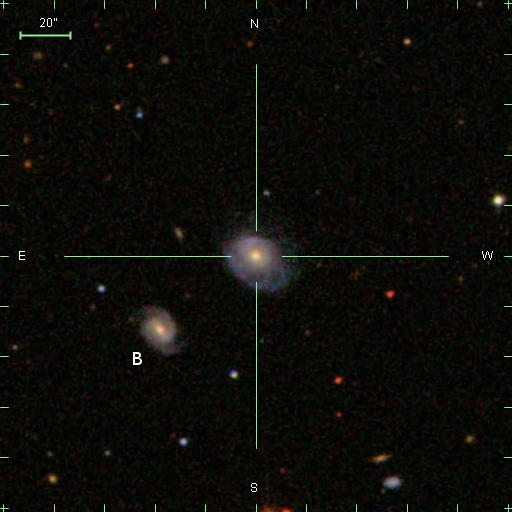}
\includegraphics[width=0.27\textwidth]{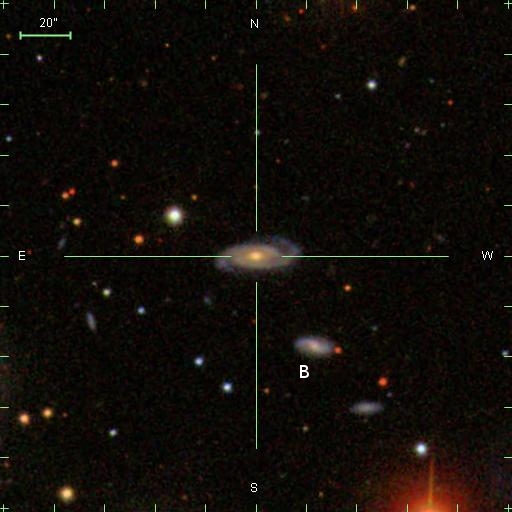}
\includegraphics[width=0.27\textwidth]{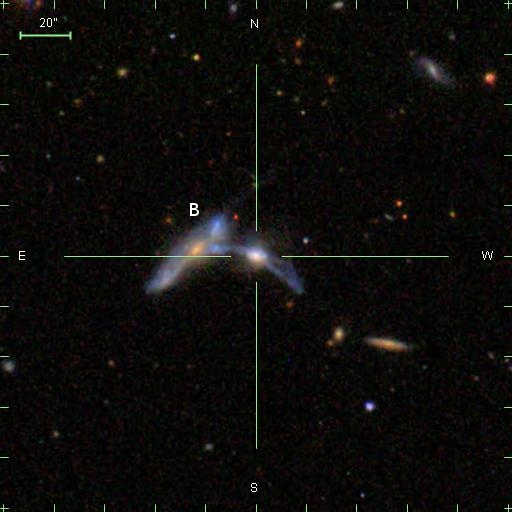} \\
\includegraphics[width=0.27\textwidth]{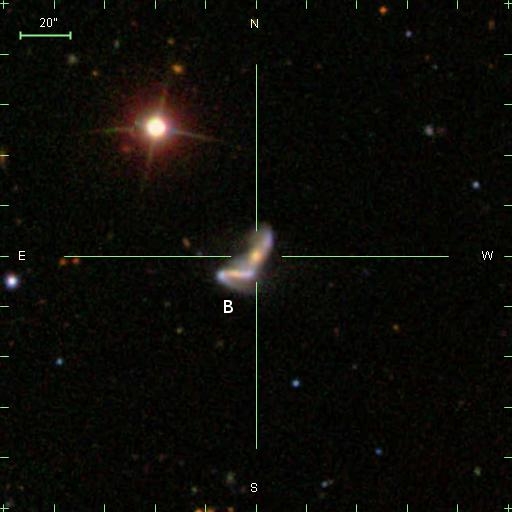}
\includegraphics[width=0.27\textwidth]{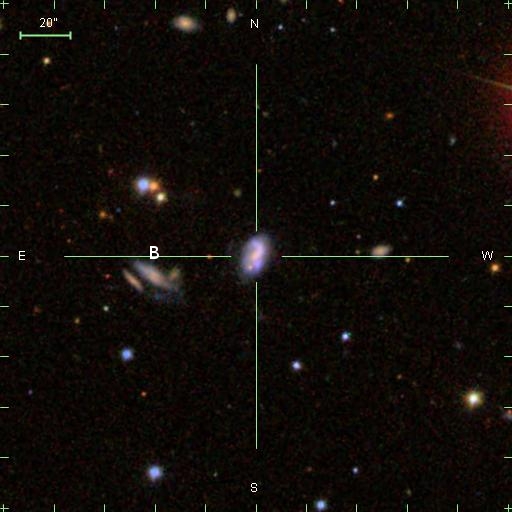}
\includegraphics[width=0.27\textwidth]{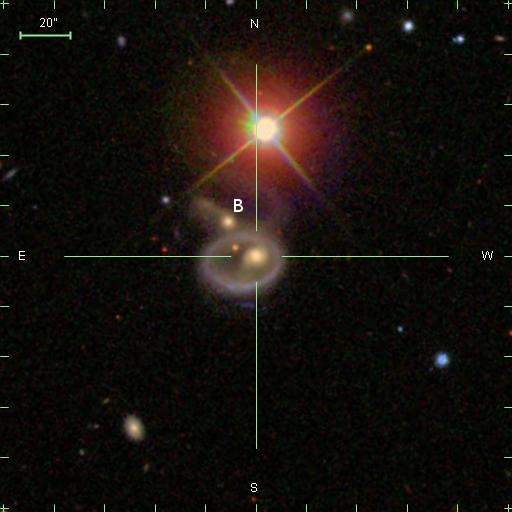}
\end{center}
\caption[Example images for isolated pairs]{SDSS three-colour images (field of view of 3\farcm4, North is up, east is to the left) of nine isolated pairs in the SDSS-DR10 footprint. The images are centred on the \emph{A} galaxy. The \emph{B} galaxy is labelled. {\it From upper left to lower right:} SIP 476, 13, 273, 153, 1079, 1123, 515, 1161, and 327.}\label{Fig:charts10pairs}
\end{figure*}

\begin{figure*}
\begin{center}
\includegraphics[width=0.27\textwidth]{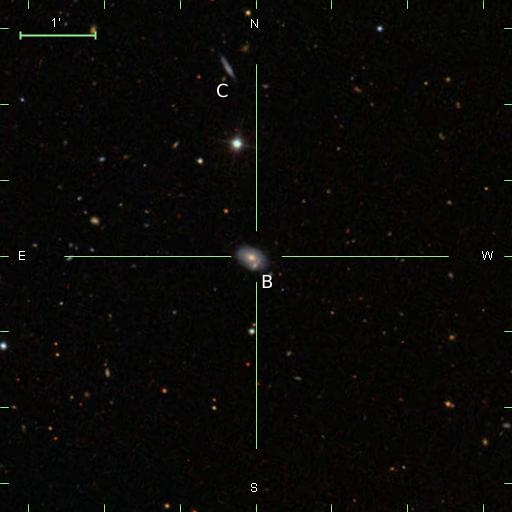}
\includegraphics[width=0.27\textwidth]{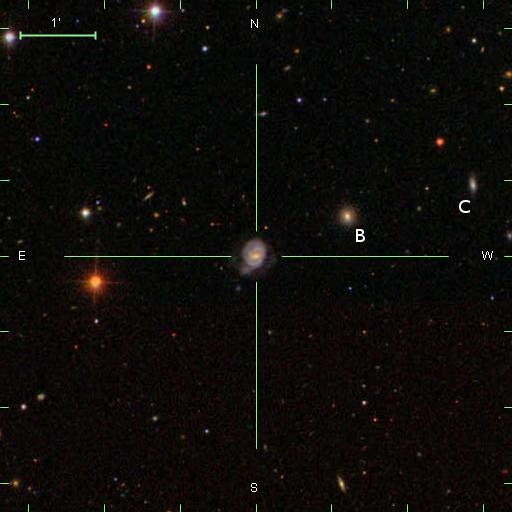}
\includegraphics[width=0.27\textwidth]{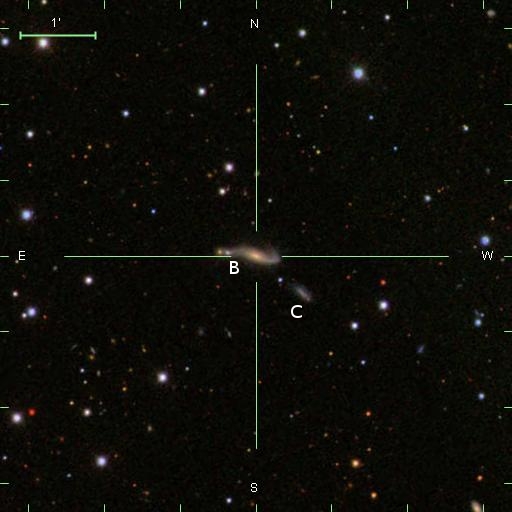}  \\
\includegraphics[width=0.27\textwidth]{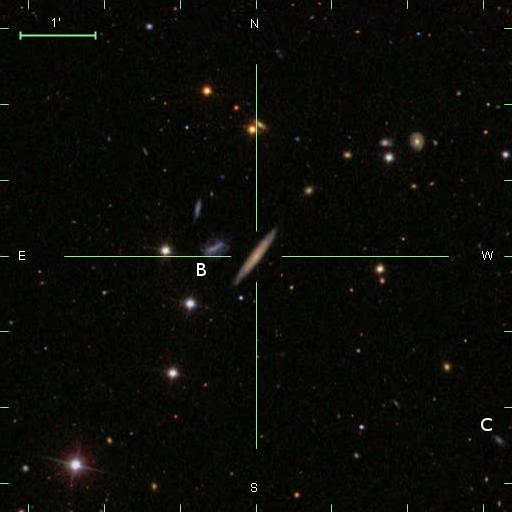}
\includegraphics[width=0.27\textwidth]{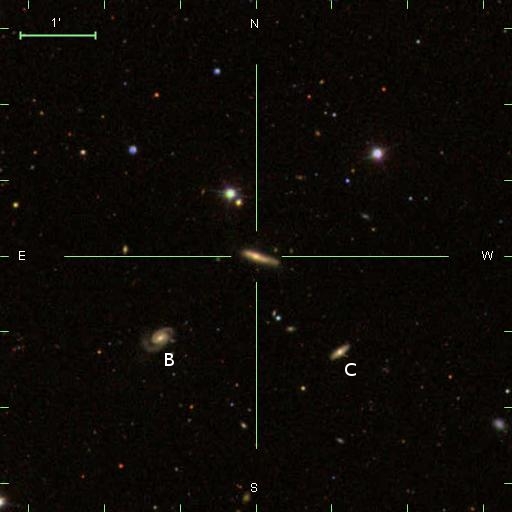}
\includegraphics[width=0.27\textwidth]{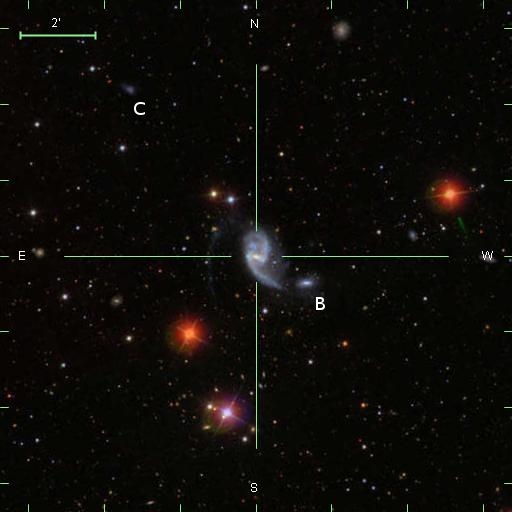} \\
\includegraphics[width=0.27\textwidth]{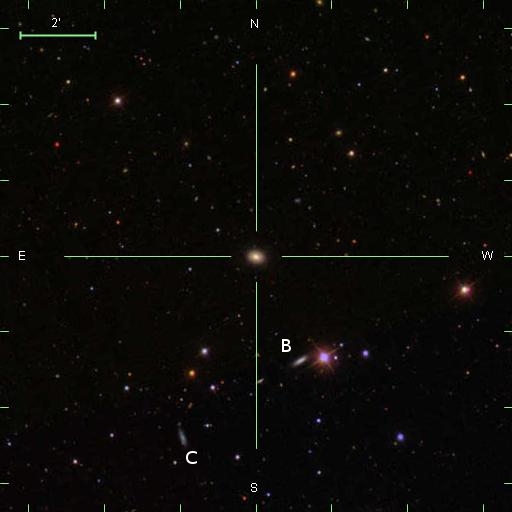}
\includegraphics[width=0.27\textwidth]{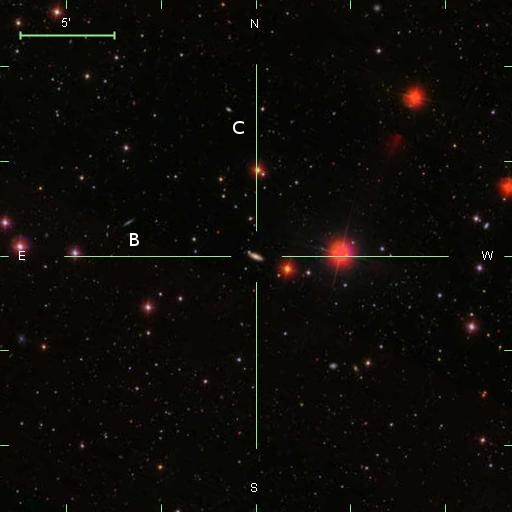}
\includegraphics[width=0.27\textwidth]{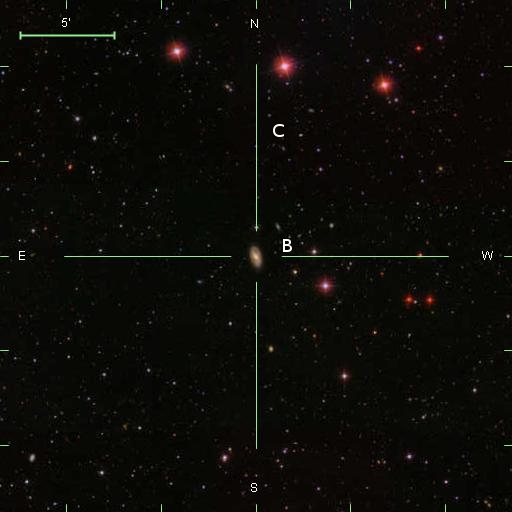}
\end{center}
\caption[Example images for isolated triplets]{SDSS three-colour images (North is up, east is to the left) of nine isolated triplets in the SDSS-DR10 footprint. The images are centred on the \emph{A} galaxy. The B and \emph{C} galaxies are labelled. {\it From upper left to lower right:} SIT 75, 12, 270, 52, and 222 with a field of view of 6\farcm7, SIT 242 and 191 with a field of view of 13\farcm5, and SIT 216 and 33 with a field of view of 27\arcmin.}\label{Fig:charts10triplets}
\end{figure*}

\end{appendix}

\end{document}